\documentclass[twocolumn,superscriptaddress,floatfix,showpacs]{revtex4-1}
\usepackage{graphics,amssymb,amsmath,epsfig,color}
\usepackage{graphicx}

\newcommand{\be}{\begin{equation}} \newcommand{\ee}{\end{equation}}
\newcommand{\bea}{\begin{eqnarray}} \newcommand{\eea}{\end{eqnarray}}

\begin{document}

\title{Information theoretic aspects of the two-dimensional Ising model}
\author{Hon Wai Lau} \affiliation{Complexity Science Group, University of Calgary, Calgary T2N 1N4, Canada}
\author{Peter Grassberger} \affiliation{Complexity Science Group, University of Calgary, Calgary T2N 1N4, Canada} \affiliation{JSC, FZ J\"ulich, D-52425 J\"ulich, Germany}
\date{\today}

\begin{abstract}

We present numerical results for various information theoretic properties of the square 
lattice Ising model.  First, using a bond propagation algorithm, we find the 
difference $2H_L(w) - H_{2L}(w)$ between entropies on cylinders of finite lengths $L$ and $2L$ with
open end cap boundaries, in the limit $L\to\infty$. This essentially quantifies how the 
finite length correction for the entropy scales with the cylinder circumference $w$. 
Secondly, using the transfer matrix, we obtain precise estimates for the 
information needed to specify the spin state on a ring encircling an infinite long cylinder. 
Combining both results we obtain the mutual information between the two halves of a cylinder 
(the ``excess entropy" for the cylinder), where we confirm with higher 
precision but for smaller systems results recently obtained by Wilms {\it et al.} -- and we 
show that the mutual information between the two halves of the {\it ring} diverges at the 
critical point logarithmically with $w$.
Finally we use the second result together with Monte Carlo simulations to show that also the 
excess entropy of a straight line of $n$ spins in an infinite lattice diverges at criticality
logarithmically with $n$. We conjecture that such logarithmic divergence happens generically 
for any one-dimensional subset of sites at any 2-dimensional second order phase transition.
Comparing straight lines on square and triangular lattices with square loops and with lines of 
thickness 2, we discuss questions of universality.

\end{abstract}

\pacs{05.50+q, 75.10.Hk, 89.70.Cf} 
\maketitle

\section{Introduction}

Although the two-dimensional Ising model is one of the best studied models and can be solved 
exactly, there are still some questions about it which are not yet settled. These concern 
in particular problems of information theoretic nature which have become important 
recently in the broader context of quantum critical phenomena, where the classical Shannon
information has is related to the von Neumann entropy, and the 
mutual entropy is related to the entanglement entropy. 

To mention just one recent result (which actually triggered the present study), it was 
shown by Wilms {\it et al.} \cite{Wilms-2011}
that the mutual information (MI) between two halves of an infinite cylinder has a maximum that 
seems to become sharper with increasing 
circumference $w$, but this maximum is not at the critical temperature but at a 
temperature $T_{\rm max} > T_c$ which does not seem to converge to $T_c$ 
when $w\to\infty$. This result, obtained by a sophisticated Monte Carlo method, is highly 
surprising, as we expect any singularity to occur only at $T_c$. One of the purposes
of the present paper is to check this by a complementary method, and to provide a simple and 
rigorous proof that the height of this maximum is $\leq 1$ bit per spin, for any $T$. 
What diverges at criticality is not the value of the MI, but its 
derivative with respect to $T$.

On the other hand, studying information 
theoretic quantities in the Ising model has a long history, with rather unclear results 
so far. The first study relevant for us was made in 1984 by R. Shaw \cite{Shaw-1984}, who 
studied the Shannon information needed to specify the spin configuration on a line 
of $n$ spins in an infinite 2-d lattice. Away from $T_c$ one expects this to be linear in $n$, 
\be
   H_n / n \to const \quad {\rm for} \;\;\; n\to\infty.
\ee
Furthermore, one expects the ``excess entropy" \cite{Shaw-1984} or ``effective measure 
complexity" \cite{Grass-1986}, defined as the MI between the 
two halves of this line as
\be
   {\cal E} = \lim_{n\to\infty} 2H_n - H_{2n},
\ee
to be finite. Due to the long range correlations at $T_c$, it is not clear whether the 
latter still holds at the critical point. Several 
simulations~\cite{Shaw-1984,Arnold-1996,Kenneway-2005,Melchert-2012} suggested 
that the excess entropy increases sharply when $T\searrow T_c$, but stays finite at $T_c$. 
A second purposes of the present paper is to show that ${\cal E}$ diverges logarithmically 
with $n$. Indeed, the MI
between the two halves of a ring encircling a cylinder show the same logarithmic divergence.
The coefficient of the logarithmic term is universal with respect to the lattice type
(square {\it vs.} triangular), but depends on the geometry of the line (straight line 
{\it vs.} topologically trivial loop on a plane lattice). It agrees numerically with the 
result obtained in \cite{Um-2012} for the ground states of quantum Ising chains.

Together with these main results, we obtain two more technical results: (i) By re-analyzing 
the numerical results of \cite{Stephan-2010} we obtain a more precise estimate 
of their universal constant $r_1$ (called $r_c$ in the present paper). And (ii) by 
obtaining transfer matrix results for widths up to $w=29$ we check the universality of 
$r_1$ with higher precision.

The rest of the paper is organized as follows: In Sec.~2 we recall some basic facts about 
mutual informations and Markov chains. In Sec.~3 we use a bond propagation algorithm (BPA) 
\cite{Loh-2006,Loh-2007,Wu-2012} to calculate the entropy of a long cylinder, from 
which we then isolate the contribution due to the open boundary conditions at its two ends.
In Sec.~4 we present results from a transfer matrix calculation and combine them with the 
results from Sec.~3 to obtain the scaling properties of the MI studied in
\cite{Wilms-2011}. We also obtain there precise estimates for $r_c$ and for the Shannon entropy 
$h_c$ per site in an infinitely long line of spins. Mutual informations between two halves of a 
ring are studied in Sec.~5. Extensive Monte Carlo simulations (using Wolff's algorithm\cite{Wolff}) 
are finally used in Sec.~6, together with the value of $h_c$ obtained in Sec.~4, to show that the 
excess entropy for such a line of $n$ spins in an infinite system 
diverges logarithmically at $T_c$. Questions of universality for this divergence are discussed
by studying also other subsets of spins that are one-dimensional in the limit $n\to\infty$.
The paper finishes with conclusions in Sec.~7. 
Several technical aspects are discussed in three appendices. 

\section{Mutual information}

In this paper we shall only deal with classical Shannon information theory \cite{Cover-Thomas}.
Given an alphabet $\Sigma=\{0,1\ldots k-1\}$ of $k$ letters and probabilities $p_i$ for 
the $i$-th letter to occur, the entropy is defined as $H=-\sum_{i=0}^{k-1} p_i \log p_i$, 
with the logarithm to base 2 if the entropy is to be measured in bits.
The following alphabets will be used in this paper:

\begin{itemize}
\item The binary alphabet $A = \{0,1\}$ or $A = \{-,+\}$, called spin $s$.
The concatenation of $n$ such spins ${\cal S}_n = (s_1 s_2 \ldots s_n)$ forms a string,
and its entropy of ${\cal S}_n$ will be denoted by $H_n$ and will be called a {\it block entropy}.

\item Each string of spin can itself be considered as a ``letter" in an alphabet
$A_n = \{-,+\}^n$ of size $2^n$. For example, the alphabet set of a spin pair is 
$A_2 = \{--,-+,+-,++\}$. We can then, just as we did above, concatenate $L$ such new 
letters to form a string, which then actually is a rectangular array of size $n\times L$.
If we have this in mind, we denote letters $\in A_n$, i.e. $n$-tuples of spins, as 
$s^{[n]} = (s_1,s_2,\ldots, s_n)$. The corresponding rectangular array is then 
denoted as ${\cal S}_L^{[n]} = (s^{[n]}_1s^{[n]}_2\ldots s^{[n]}_L)$.

\item In particular we shall consider strings of length $n=w$, where $w$ is the 
width of the lattice. 
We will assume that the lattice is periodic laterally, i.e. $w$ is actually the 
circumference of a cylinder, and $s^{[w]}$ can be viewed as forming a ring.
The entropy of $L$ adjacent such rings, i.e. of a rectangular $w\times L$ array with periodic 
b.c. in the $w$-direction and open b.c. in the $L$-direction will be 
called $H_L(w)$. 
\end{itemize}

For any joint probability distribution over a product of alphabets $\cal{A}$ and $\cal{B}$
the MI is defined as \cite{Cover-Thomas}
\bea
   I({\cal A:B}) & = & H({\cal A})+H({\cal B})-H({\cal AB}) = H({\cal A})-H({\cal A}|{\cal B}) \nonumber \\
                & = & \sum_{i\in {\cal A}, j\in {\cal B}} p_{ij} \log\frac{p_{ij}}{p_ip_j} \;.    \label{I}
\eea
It is equal to the average decrease in code length needed to specify the value $i$ in a random
realization, if the value of $j$ gets known and if encoding is done optimally for many such 
independent realizations jointly.

If $\cal{A}$ in Eq.(\ref{I}) is a set of length-$n$ strings and $\cal{B}$ the set of single letters,
then 
\be
   h_n \equiv H({\cal AB}|{\cal A}) = H({\cal S}_{n+1})-H({\cal S}_n) = H_{n+1}-H_n
\ee
(with $H_0=0$) is the information needed to specify the last one in a string of $n+1$ letters, 
given all previous ones. If the source generating the string is ergodic and the limit exists, then 
\be
   h = \lim_{n\to\infty} h_n
\ee
is called the entropy per letter of the source, or simply the entropy per letter.
If, moreover, the probability distribution is stationary then $h_n$ is monotonically decreasing with
$n$, since the difference 
\be
   \delta h_n = h_{n-1}-h_n 
\ee
can be interpreted as the amount by which the uncertainty about the last of $n+1$ letters decreases, 
if the first one gets known (and all intermediate ones are known already) \cite{Grass-1986}. An 
alternative interpretation of $\delta h_n$ is as a conditional mutual information 
\cite{Cover-Thomas}, $\delta h_n = I(s_{n+1}:s_1|s_2\ldots s_n)$.

Of particular importance are Markov chains. A Markov chain of order $k$ is characterized by 
\be
   p_{s_n|s_{n-1},s_{n-2},s_{n-3},\ldots} = p_{s_n|s_{n-1},\ldots s_{n-k}}\;,
\ee
i.e. the memory is of length $\le k$. Notice that this does not imply that there are no longer 
ranging correlations, but they are all mediated by a chain of short range steps. For a Markov 
chain of order $k$ it is easily seen that \cite{Grass-1986}
\be
   \delta h_n = 0 \quad {\rm for} \;\;\; n>k\;.
\ee
Thus for a first order Markov chain $\delta h_1 = 2H_1-H_2 \ge 0$, while $\delta h_n = 0$ 
for $n \ge 2$. 

Notice that this notation assumes that the chain is in its stationary state,
i.e. $H_1$ does not refer to the entropy of the first letter(s), if there is a transient. But the 
basic result is more general. Consider e.g. a heterogeneous and non-stationary first-order Markov 
chain ${\cal A} \text{--} {\cal B} \text{--} {\cal C} \text{--} {\cal D}$. Then 
\be
  I({\cal AB}:{\cal CD}) = I({\cal B}:{\cal C}). \label{II}
\ee
This allows an immediate generalization to Markov fields. A (first-order) Markov field is a graph
with random variables at the vertices, such that any two subsets of nodes ${\cal A, C}$ become
independent, if one conditions on a {\it separating set} ${\cal B}$ (the set ${\cal B}$ separates
${\cal A}$ and ${\cal C}$, if every path connecting the latter has to pass through  ${\cal B}$).
Consider now a splitting of the entire graph into two disjoint subsets ${\cal A, C}$. Furthermore, 
divide each subsets into its {\it interior} and its {\it boundary} , where the latter is the set of 
nodes with links to the other subset. This defines then a Markov chain 
${\cal A}_0 \text{--} \partial{\cal A} \text{--} \partial{\cal C} \text{--} {\cal C}_0$,
where the subscript ``0" indicates the 
interior and ``$\partial$"  indicates the boundary. Eq.~(\ref{II}) gives then that the MI
between any two subsets is equal to the MI between their boundaries,
\be
  I({\cal A}:{\cal C}) = I(\partial{\cal A}:\partial{\cal C}). \label{dII}
\ee
and thus smaller than the entropy of either boundary.

For a bi-infinite string $S = (\ldots s_{-1}s_0s_1\ldots)$ the {\it excess entropy} 
\cite{Shaw-1984} or {\it effective measure complexity} \cite{Grass-1986} is defined as the MI 
between the left half $S^{-} = (\ldots s_{-1}s_0)$ and the right half $S^{+} = (s_1s_2\ldots)$
or, equivalently, as 
\be
   {\cal E}(S) = \sum_{n=0}^\infty (h_n - h) = \sum_{n=1}^\infty \delta h_n\;.
\ee
For a first-order Markov chain we have simply
\be
   {\cal E}(S) = \delta h_1\;,
\ee
i.e. the MI between left and right halves is just equal to the MI
between two neighboring letters.

A first application of this to the 2-d Ising model on an infinitely long strip of width $w$ (with  
either periodic or open lateral boundary condition) 
is that the excess entropy per spin is equal to the MI between
two adjacent lines (resp. rings) of $w$ spins, and is thus bounded by $w$ bits, because the 
transfer matrix generates a first order Markov process. This explains immediately why 
the MI per width measured in \cite{Wilms-2011} is finite for all $T$, even in the limit
$w\to\infty$ and $T\to T_c$. In order to compute this excess entropy explicitly, we need 
two ingredients:
Because $\delta h_1 = h_1 - h_0$ , we need both the unconditional information 
$h_0 = H_w$ for a ring and the information $h_1 = H_2(w)-H_w$ for a ring conditioned on 
one of its neighbors. The first one will be computed by the transfer matrix (TM), and the 
second by the bond propagation algorithm (BPA) 
We will show that the combination of both algorithms allows to obtain mutual informations
for up to $w \approx 30$, about twice the size feasible on a workstation 
with the TM alone.

\section{Cylinder entropies obtained by bond propagation}

The bond propagation algorithm (BPA) for the Ising model \cite{Loh-2006,Loh-2007} is a 
modification of a similar algorithm developed by Frank and Lobb \cite{Frank-1988} for
finding the resistance of a 2--d resistor network. In contrast to transfer matrix methods
it cannot be used to obtain probabilities of spin configurations (such as those needed to
calculate the entropy of a single ring), but it is the most efficient and accurate method
known so far for calculating free and internal energies of finite 2-d systems 
\cite{Wu-2012}. It was used until now only for open lateral b.c., but as pointed out in 
\cite{Wu-2012} it can also be adapted to periodic b.c. in one direction (but not in 
both). The basic strategy is described in 
\cite{Frank-1988}, and details are given in \cite{Loh-2006,Loh-2007,Wu-2009}. 

We implemented the BPA for the Ising model with cylindrical b.c. with sizes $L \times w$.
Although we are interested in the limit
$L/w\to\infty$, we found that $L = 10w$ is in general enough to obtain results precise up to 
machine precision. For $w<20$ we checked our results also by means of a microcanonical
transfer matrix \cite{Binder-1972,Creswick-1995} that is very accurate in our implementation.

In the limit $L,w\to \infty$ one obtains of course Onsager's 
result \cite{Onsager-1944} which reads, at inverse 
temperature $\beta=\beta_c =\ln(1+\sqrt{2})/2 = 0.44068679\ldots$, 
\be
   h_c := \lim_{L,w\to\infty} \frac{H_L(w,\beta_c)}{Lw} = 0.442142977\ldots {\rm bits}\;.    \label{hw}
\ee

The values 
\be
   \tilde{h}(w,\beta)=\lim_{L\to\infty}\frac{H_{L}(w,\beta)}{Lw}
\ee
indeed converge at $\beta=\beta_c$ to $h$ as
\be
   {\tilde h}(w,\beta_c) \approx h_c-\frac{0.2618}{w^2}+\frac{0.15}{w^4}+\frac{0.5}{w^6} +\ldots\;.
\ee

This conforms with the result of \cite{Izmailian-2001,Izmailian-2002} that the free and internal 
energies per site are power laws in $1/w^2$.

\begin{figure}
\includegraphics[width=0.5\textwidth]{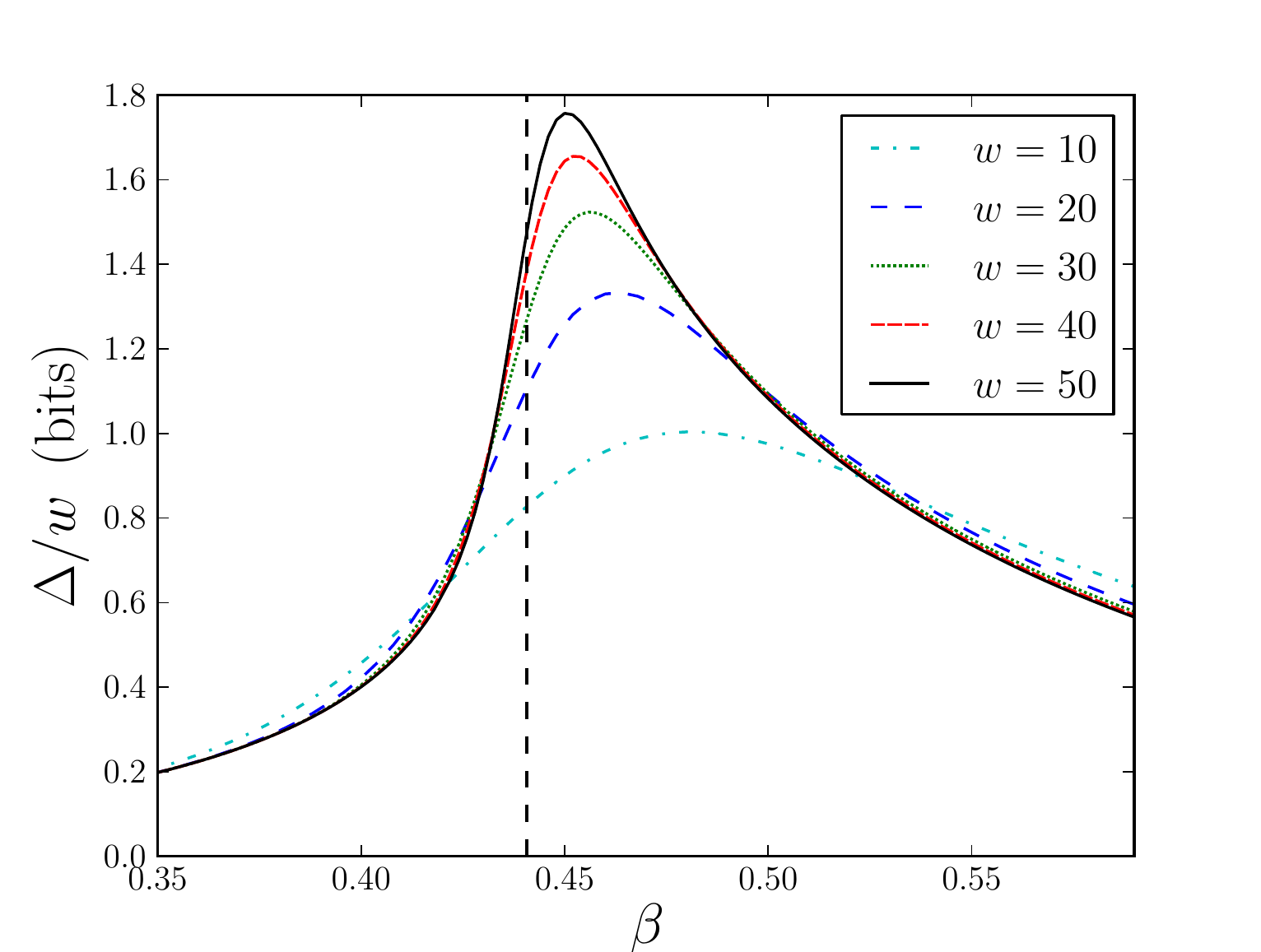}
\caption{(Color online) The quantity $\Delta(w,\beta)$ defined in Eq.~(\ref{Dhw}),
for critical Ising systems of finite widths $w$, plotted against $\beta$.
The vertical dotted line is the critical $\beta_c$.}
   \label{fig1}
\end{figure}

More interesting for us, however, is the detailed convergence with $L$. We assume 
that the limit $\lim_{L\to\infty} (H_L(w,\beta)/w - {\tilde h}(w,\beta)L)$ exists. 
In this case we can calculate it by comparing two cylinders of length $L$ to one cylinder 
of length $2L$, and obtain
\be
   \Delta(w,\beta) = \lim_{L\to\infty} [2H_L(w,\beta)-H_{2L}(w,\beta)].     \label{Dhw}
\ee
We found that this limit indeed converged very rapidly. Away from the critical region $\Delta(w,\beta)/w$
also stays bounded for $w\to\infty$, but not near $\beta=\beta_c$, as seen from Fig.~\ref{fig1}.

\begin{figure}
\includegraphics[width=0.5\textwidth]{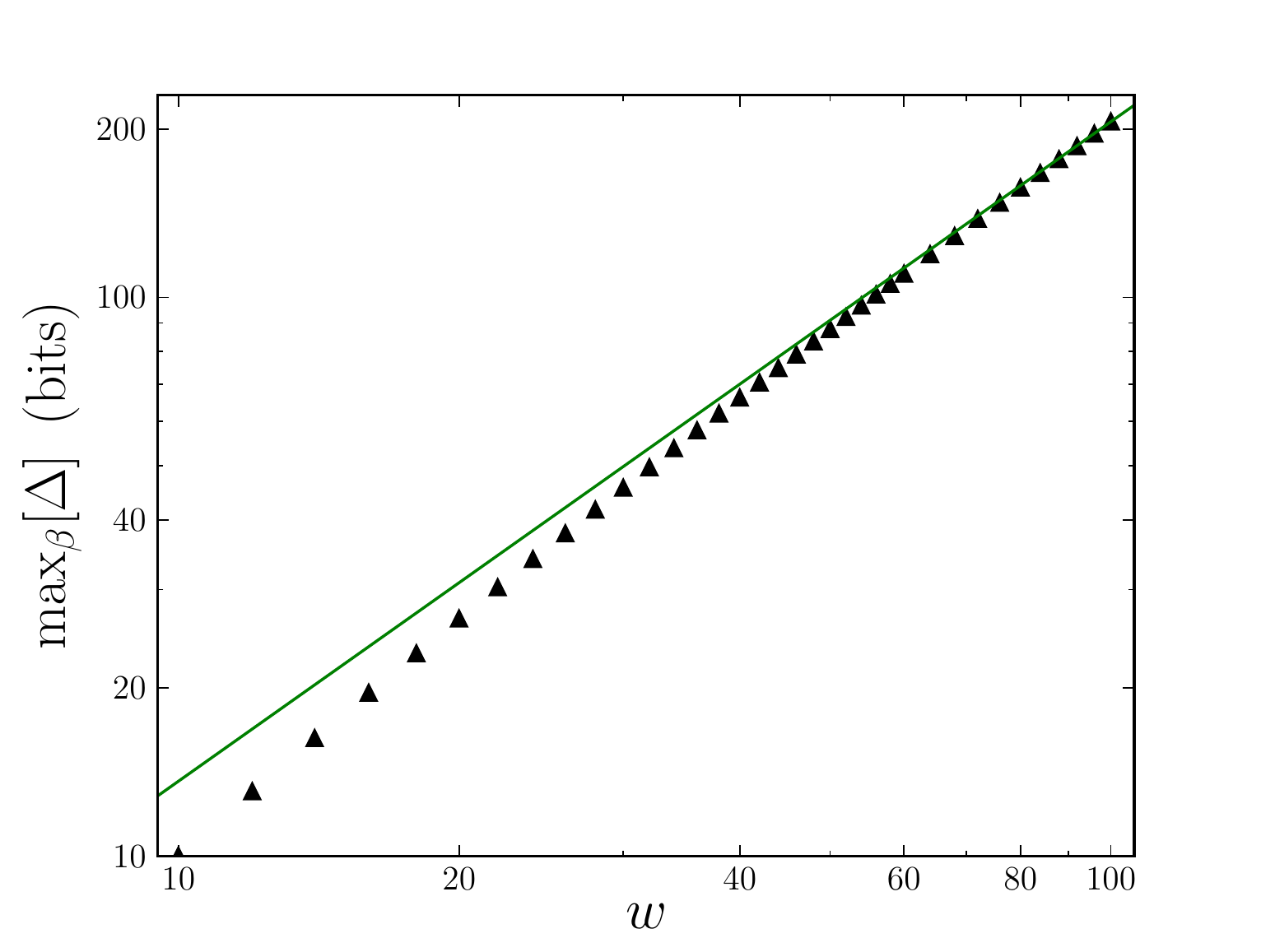}
\caption{(Color online) Log-log plot of the peak heights of Fig.~\ref{fig1}, i.e. of the maximal 
values of $\Delta(w,\beta)$, versus $w$. The straight line has slope $1.18$. }
   \label{fig2}
\end{figure}

\begin{figure}
\includegraphics[width=0.5\textwidth]{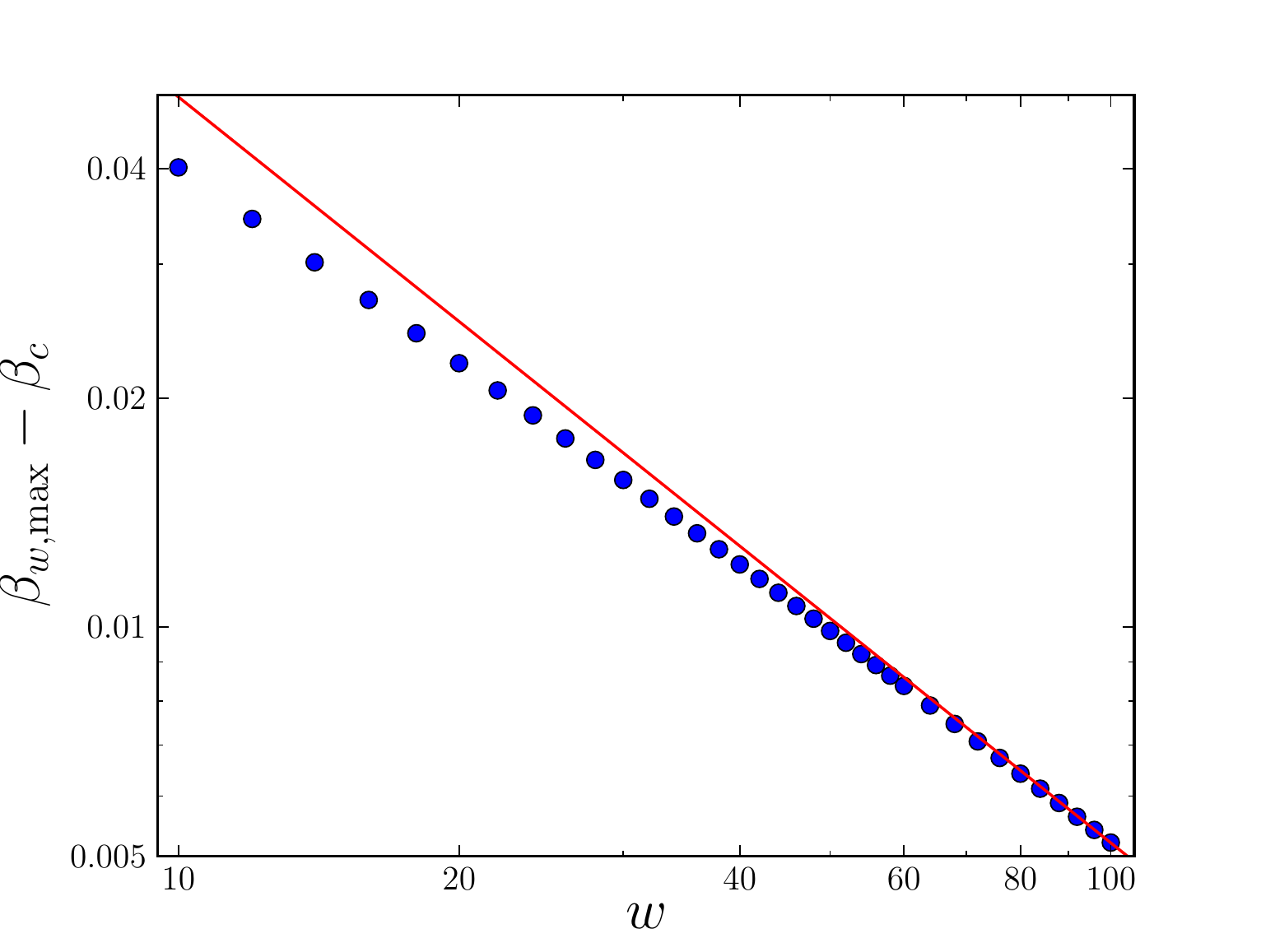}
\caption{(Color online) Log-log plot showing the scaling of the peak positions of Fig.~\ref{fig1}, 
versus $w$. The straight line has slope $-0.98$, indicating that 
$\beta_{w,\rm max} -\beta_c \sim w^{-0.98}$. }
   \label{fig3}
\end{figure}

A more detailed study shows that both the peak height in Fig.~\ref{fig1} and the distance of the 
peak position from $\beta_c$ scale like powers of $w$ (see Figs.~\ref{fig2},\ref{fig3}),
\be
   \max_{\beta}[\Delta(w,\beta)] \sim w^{1.18\pm 0.03}\;,
\ee
\be 
   \beta_{w,\rm max} -\beta_c \equiv \underset{\beta}{\arg\max} [\Delta(w,\beta)] -\beta_c \sim w^{-0.98\pm 0.03}\;.
\ee
The errors are here rather large due to large corrections to scaling.

From an information theoretic point of view, $\Delta (w,\beta)/w$ has two contributions. On the one hand,
the system consisting of two independent cylinders of length $L$ misses all horizontal bonds in 
the center of the cylinder of length $2L$ which connect the left and right halves. This 
contributes a term which basically 
measures the effect of the free end boundary condition as compared to,
say, periodic boundary conditions. But this is not the only effect. Even if these bonds were present,
the information needed to specify the two cases would differ by the MI between the 
left and right halves, i.e. by the excess entropy. As we have already pointed out in Sec.~2, this 
excess entropy (per unit of $w$) is always bounded, so that the first contribution alone is 
responsible for the divergence.

\section{Shannon Information of a Single Ring and Excess Entropy of a Cylinder}

Since any transfer matrix of the Ising model on a finite width strip 
induces a Markov chain, the entropy $w{\tilde h}(w,\beta)$ per ring studied in the previous section 
is just the conditional Shannon entropy of this ring, conditioned on the previous one, i.e. in 
the notation of Sec.~2 
\be
   {\tilde h}(w,\beta) = (H_2(w,\beta) - H_1(w,\beta))/w 
\ee
(notice again that we assume stationarity, i.e. $H_1(w,\beta)$ refers to a single 
ring far from the ends of the cylinder).
To skip the transient state, this quantity can be calculated by 
$w{\tilde h}(w,\beta) = \lim_{L\to \infty} [H_{L+1}(w,\beta) - H_L(w,\beta)]$ using 
the results obtained with the BPA in Sec.~3.

In order to obtain the excess entropy ${\cal E}(w) := \lim_{L\to\infty} 
{\cal E}({\cal S}_L^{[w]})=2H_1(w)-H_2(w)$, or mutual information in this case, 
of the infinitely long cylinder, we need in addition values of $H_1(w)$, i.e. of the unconditioned
Shannon entropy of this line. This is obtained from a conventional transfer matrix (TM) calculation.
Details of this calculation are given in Appendix~A. At $\beta=\beta_c$ we obtained data for $w$
up to 29, at a few selected points away from criticality up to $w=28$.

In principle one can obtain in this way also the Shannon entropy $H_2$ for two adjacent rings.
This would, however, require to estimate all probabilities over $2^{2w}$ spin configurations.
With present workstations this can be done only for $w\lesssim 16$. On the other hand, the BPA
or a similar scheme can find the entropy of the whole lattice up to much larger $w$ -- but it 
cannot give the entropy of a single ring. By using the Markov chain property and combining
the BPA with the TM, we can therefore compute the exact numerical excess entropy for $w$
up to $\approx 30$.

\begin{figure}
\includegraphics[width=0.5\textwidth]{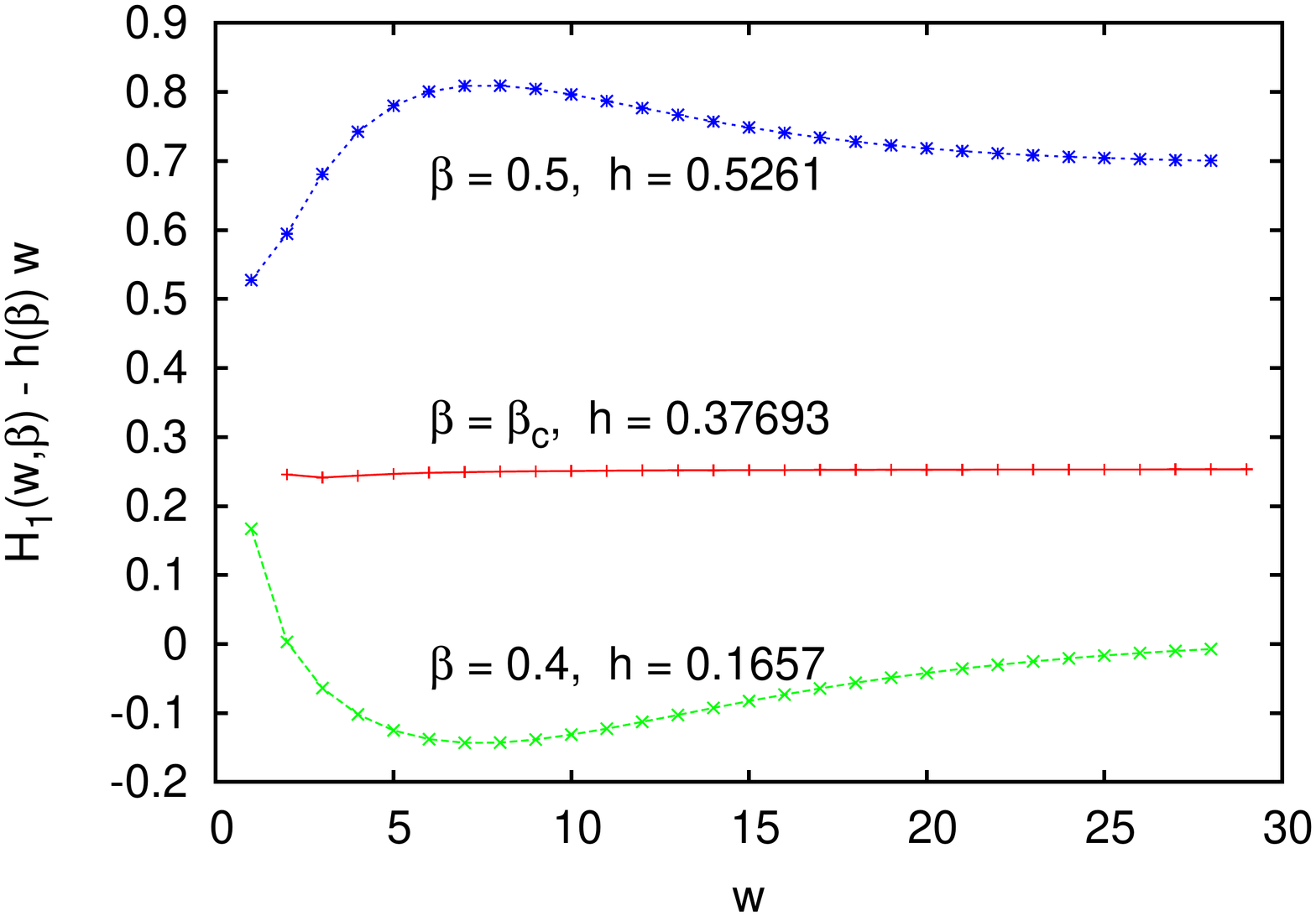}
\caption{(Color online) Shannon entropies (in nats) for a ring encircling a cylinder of width $w$,
after subtracting the leading term $\propto w$. }
\label{stephan1.fig}
\end{figure}

We first checked that our TM data were indeed consistent 
with the conjecture of St\'ephan {\it et al.} \cite{Stephan-2010, stephan_shannon_2009}
\be
   H_1(w,\beta) = h(\beta)w+r(\beta)+o(1)
\ee
for $w\to\infty$, where
\be
   r(\beta) = \left\{ \begin{array}{rcl}
                         0 & \mbox{for} & \beta < \beta_c \\
         r_c:=0.2543925(5) & \mbox{for} & \beta = \beta_c \\
                   \ln(2) & \mbox{for} & \beta > \beta_c
                      \end{array}\right.                     \label{stephan}
\ee   
while $h(\beta)$ is the Shannon entropy per spin for an infinitely wide cylinder.
For this we plot in Fig.~\ref{stephan1.fig} the differences $H_1(w,\beta) - h(\beta)w$ for three 
values of $\beta$ against $w$, where $h(\beta)$ is chosen such that the curves become flat 
for $w\to\infty$. We find perfect agreement. Moreover we see that the leading corrections 
for large $w$ are $\sim 1/w$ for $\beta = \beta_c$, $\sim 1/w^2$ for $\beta < \beta_c$, and 
$\sim 1/w^3$ for $\beta > \beta_c$. 

Although Fig.~\ref{stephan1.fig} is sufficient to verify Eq.(\ref{stephan}), it does not do justice
to the very high precision of the transfer matrix data of \cite{Stephan-2010}. In particular, we 
want for Sec.~5 a much more precise estimate of the entropy per spin for the critical case. 
Therefore we first re-analyzed the data of \cite{Stephan-2010} to obtain a more precise estimate
\be
   r_c=0.254392505(10)\; {\rm nats} = 0.367010805(14)\; {\rm bits}.    \label{rc}
\ee
Details are given in Appendix B. After that, universality is invoked to use this value
as a constraint in a similar analysis, in order to obtain
\be
   h(\beta_c) = 0.37692626(7)\; {\rm nats} = 0.54378965(10) \;{\rm bits}.   \label{hc}
\ee
Again details are given in Appendix B.

\begin{figure}
\includegraphics[width=0.5\textwidth]{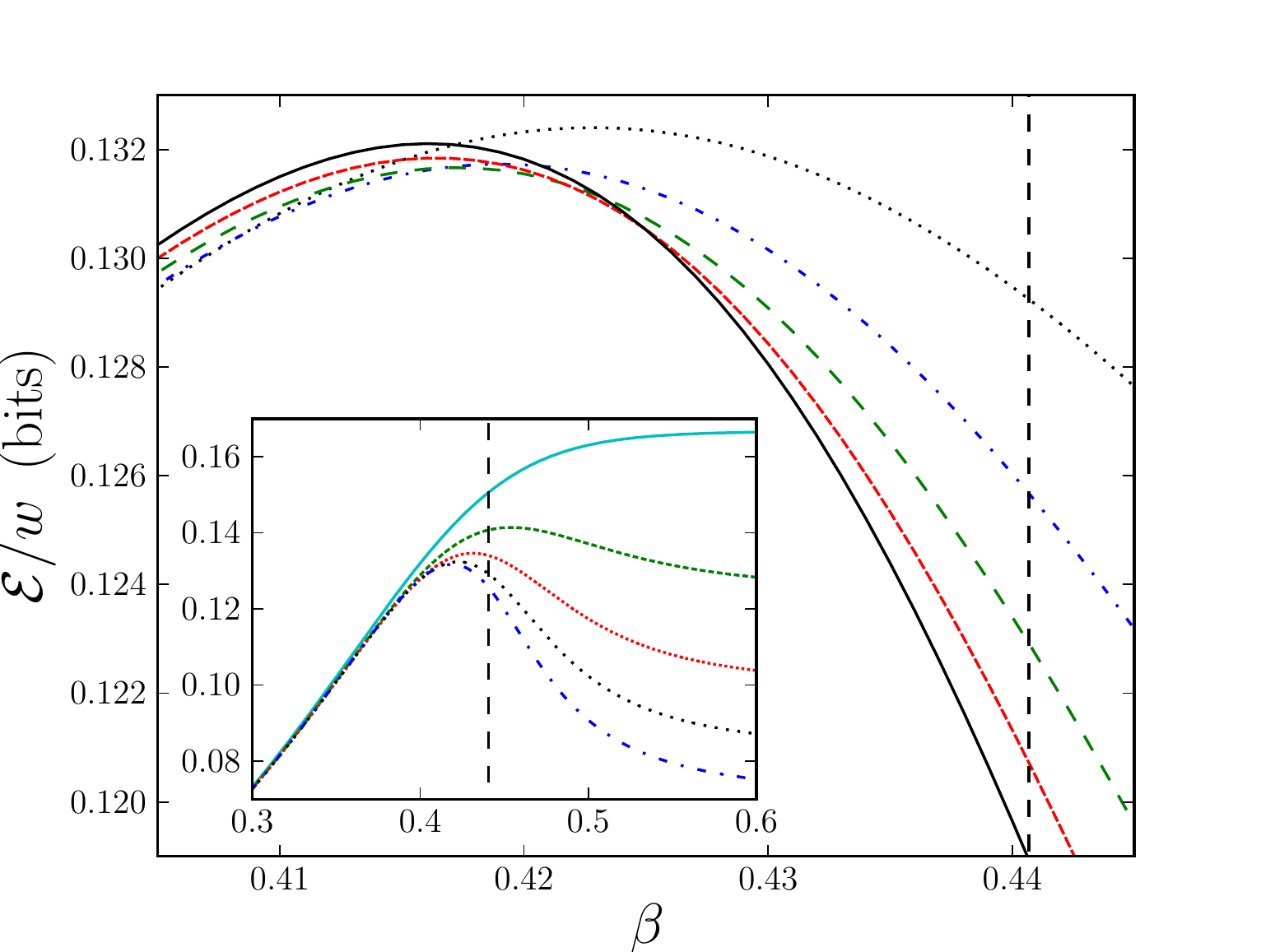}
\caption{(Color online) Excess entropy ${\cal E}(w,\beta)$ (in bits per unit width) between two halves of a 
cylinder. From top to bottom (at the critical temperature), the curves correspond to $w=12,14,16,18,20$. 
    The inset shows the golbal behavior for $w=6,8,10,12,14$. For $w<6$ the curves are monotonically increasing.}
   \label{fig.Mperw}
\end{figure}

\begin{figure}
\includegraphics[width=0.5\textwidth]{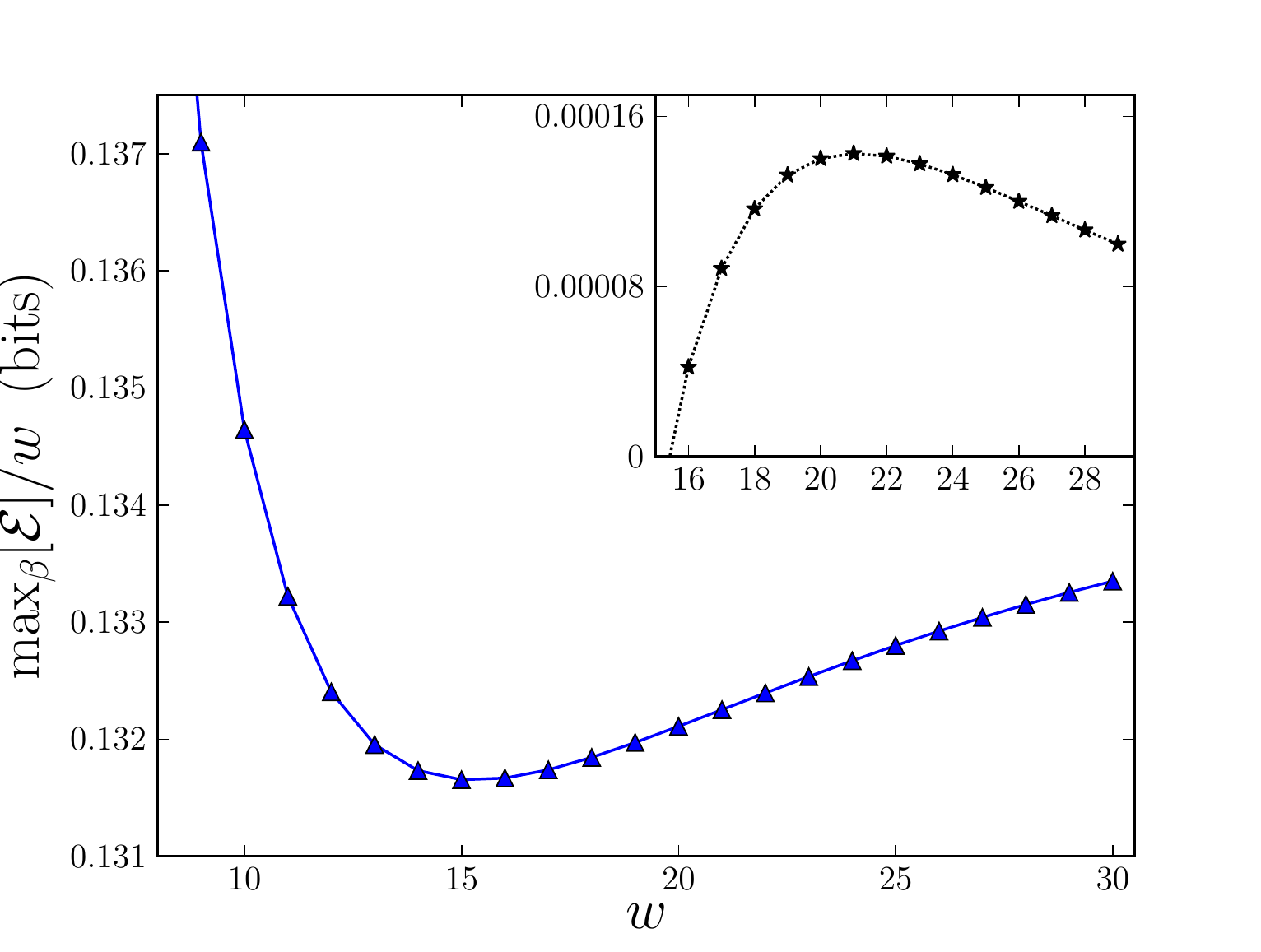}
\caption{(Color online) Peak heights (in bits) of the excess entropy in Fig.~\ref{fig.Mperw}.
A minimum and an inflection point are located at $w=15$ and $w=21$ respectively.
The inset shows the derivative of the height with respect to $w$.
}
   \label{fig.Mperw_height}
\end{figure}

\begin{figure}
\includegraphics[width=0.5\textwidth]{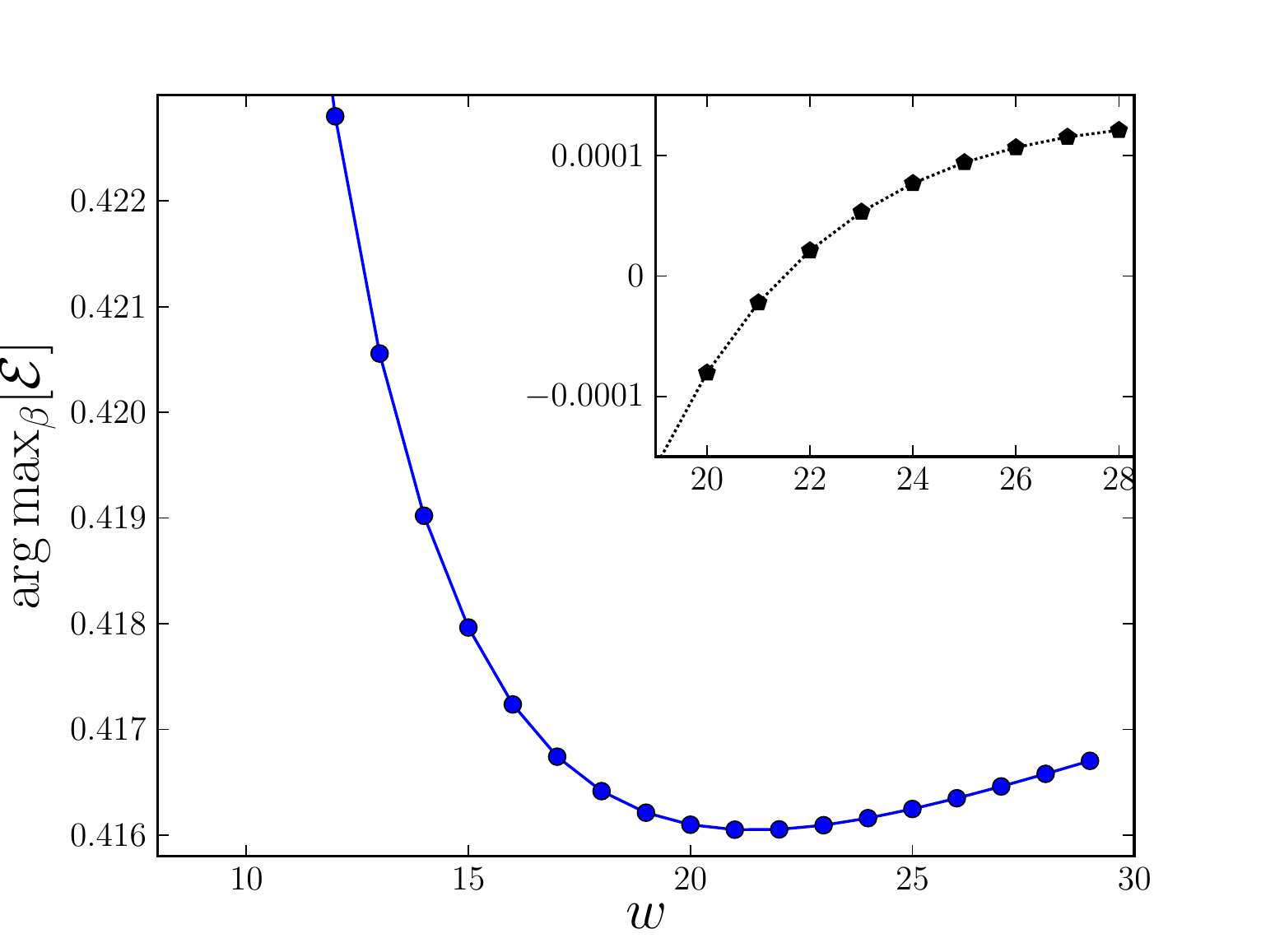}
\caption{(Color online)
Peak positions of the excess entropy in Fig.~\ref{fig.Mperw}.
Again the inset shows the slope of the curve.}
   \label{fig.Mperw_location}
\end{figure}

After having verified the correctness of the algorithm, we now proceed to obtain the 
excess entropy ${\cal E}(w,\beta) = H_1(w,\beta)-w\tilde{h}(w,\beta)$.
The data is given in Fig.~\ref{fig.Mperw}. The overall picture is shown in the inset,
while the main figure shows an enlargement close to the critical region.
We indeed verify the qualitative behavior found in \cite{Wilms-2011}.
In particular, we find that the curves for $w > 8$ have peaks in the region $\beta<\beta_c$, 
and that the peak positions shift to smaller values of $\beta$ as $w$ is increased.
As seen from Fig.~\ref{fig.Mperw_height}, the peak heights first decrease with $w$
and reach a minimum at $w=15$. They then increase again, but the rate of increase
slows down for $w>21$ (see inset of Fig.~\ref{fig.Mperw_height}). The peak positions 
(Fig.~\ref{fig.Mperw_location}) show a similar behaviour, and reach a minimum at $w=21$.
Extrapolating these data to $w=\infty$ is not easy, but our best estimates are 0.417(2)
for the position and 0.134(2) for the height. 
Both are consistent with the less precise results of Wilms {\it et al.} 
\cite{Wilms-2011} obtained for larger systems. While the peaks initially get sharper with 
increasing $w$, their widths soon seem to become independent of $w$.
All this indicates that these peaks have no close relationship with criticality.

\begin{figure}
\includegraphics[width=0.5\textwidth]{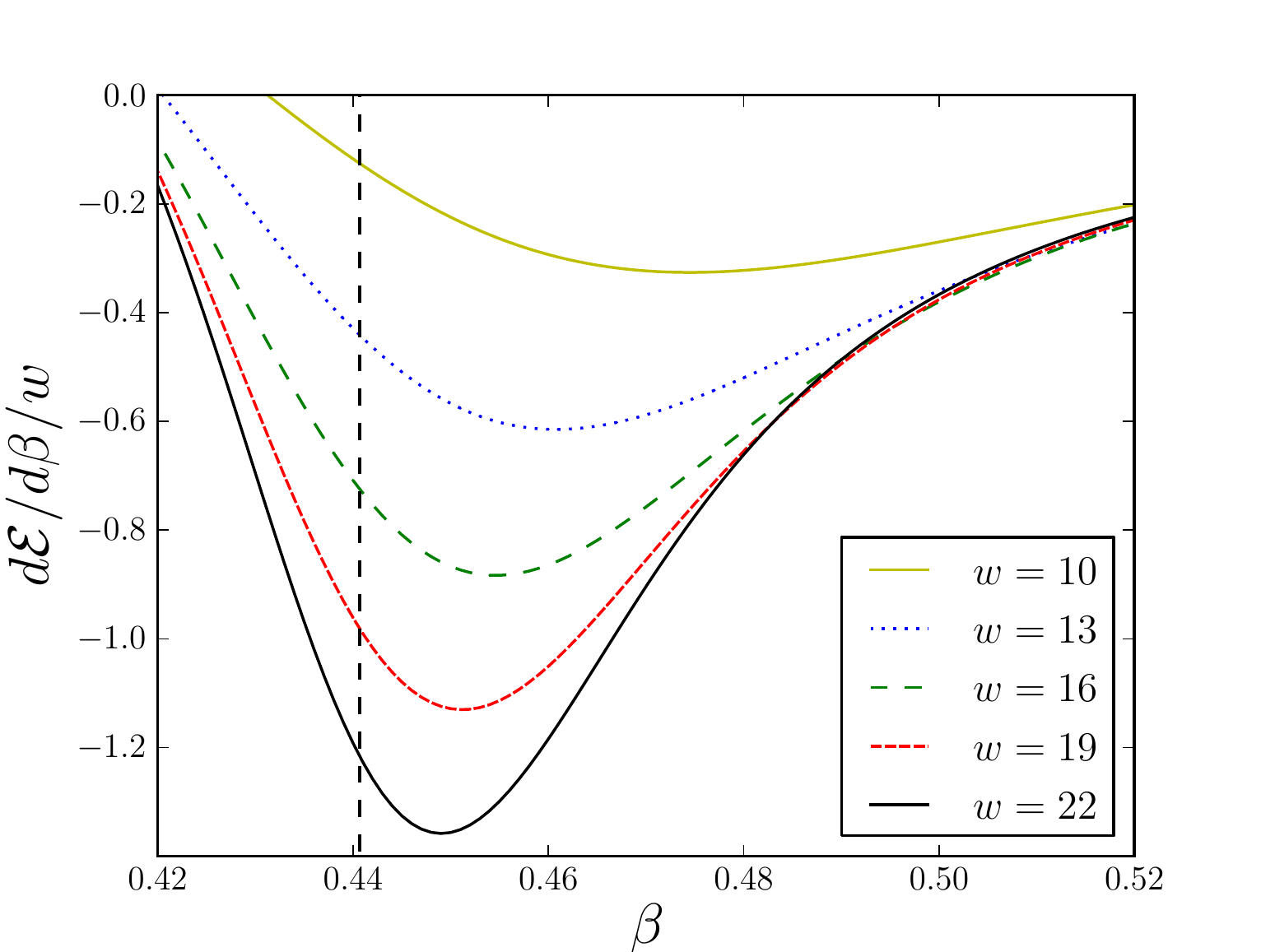}
\caption{(Color online) Slope of the excess entropy $d{\cal E}(w,\beta)/d\beta$ per width in
    Fig.~\ref{fig.Mperw}.}
   \label{fig.dMperwdbeta}
\end{figure}

\begin{figure}
\includegraphics[width=0.5\textwidth]{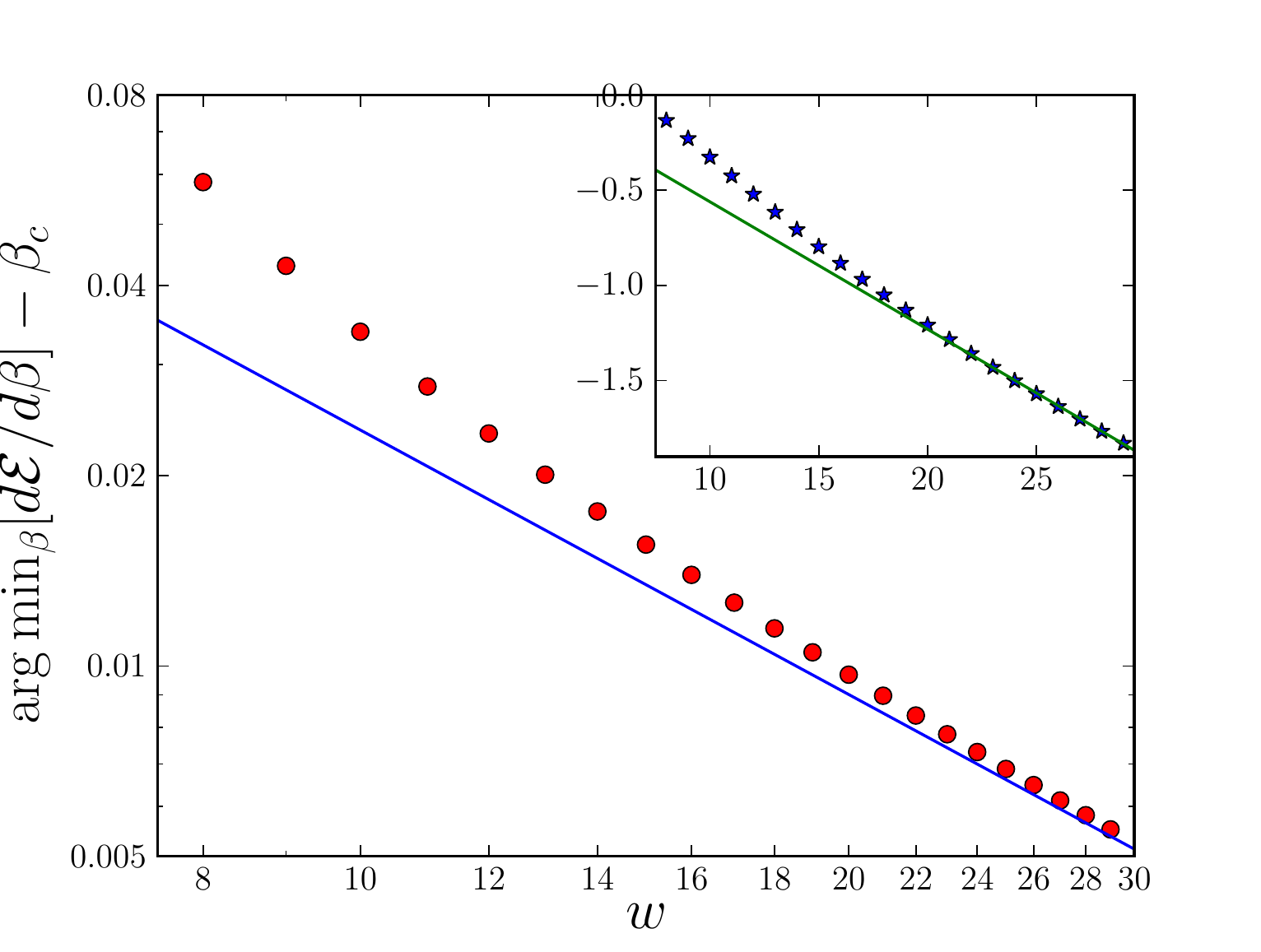}
\caption{(Color online) Log-log plot of the positions of the minimum in Fig.~\ref{fig.dMperwdbeta}.
The straight line has slope -1.39.
The inset shows the value of the minimum in Fig.~\ref{fig.dMperwdbeta} on a linear plot.
The straight line indicates a power }
   \label{fig.dMperwdbeta_min}
\end{figure}

On the other hand, the slopes (obtained by numerical differentiation) of the 
mutual information in Fig.~\ref{fig.Mperw} seem to diverge to $-\infty$ 
near $\beta_c$, as seen from Fig.~\ref{fig.dMperwdbeta}.
A more detailed analysis (Fig.~\ref{fig.dMperwdbeta_min}) shows that the 
minimum of the slope (i.e. the inflection point) moves towards $\beta_c$ 
according to a power law $\beta_{w,{\rm min}}-\beta_c \sim w^{-1.39 \pm 0.08}$,
where $\beta_{w,{\rm min}} = \arg \min_\beta [d{\cal E}(w,\beta)/d\beta]$.
Moreover, a linear fit of the minimum value is also shown in the figure, 
suggesting a power law 
\be
   \frac{1}{w}\min_\beta [d{\cal E}(w,\beta)/d\beta] \sim \frac{1}{w^\alpha}
\ee
where $\alpha$ equal to or slightly less than 1. 
The uncertainties in both exponents are large due to large corrections to scaling.
All this suggests strongly that it is the slope of the mutual information, not the 
mutual information itself, that diverge at the critical point.

\section{Mutual information between parts of a single ring}

The fact that the MI between the two halves of a cylinder does not diverge, demonstrated in the 
previous section, does not tell anything about MIs between parts of the ring separating the cylinder
halves. Let us divide a ring of length $w$ into two halves of lengths $m$ and $w-m$. The transfer
matrix calculations of the previous section allow us also to compute the Shannon entropies of these 
parts, and hence of the MI between them.

\begin{figure}
\includegraphics[width=0.5\textwidth]{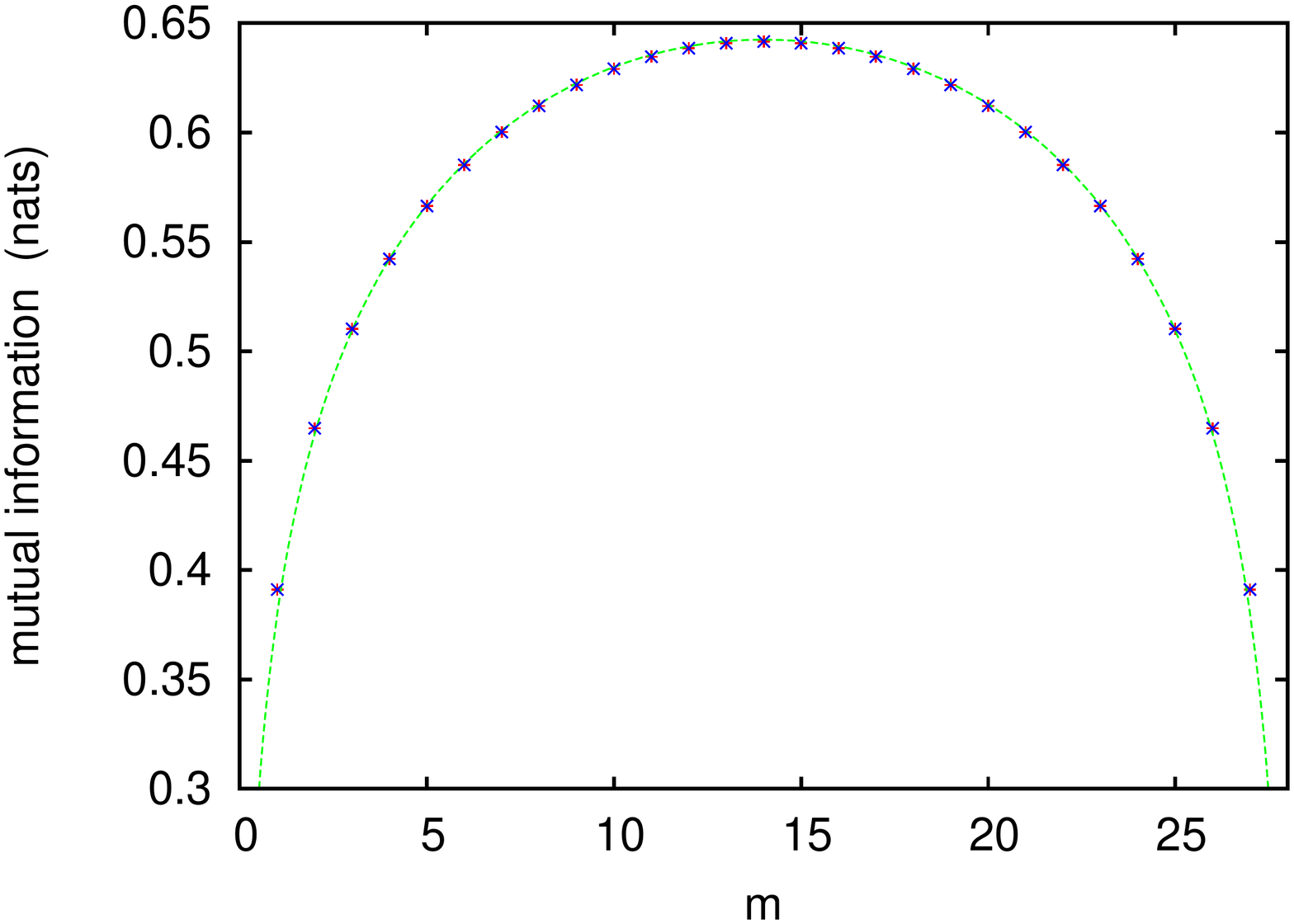}
\caption{(Color online) Mutual informations between two parts of a ring of $w=28$ spins encircling a 
cylinder (in natural units), plotted against the length $m$ of one of the parts. The continuous 
line corresponds to Eq.~(\ref{ring-MI2-fit}).}
 \label{ring-MI2.fig}
\end{figure}

\begin{figure}
\includegraphics[width=0.5\textwidth]{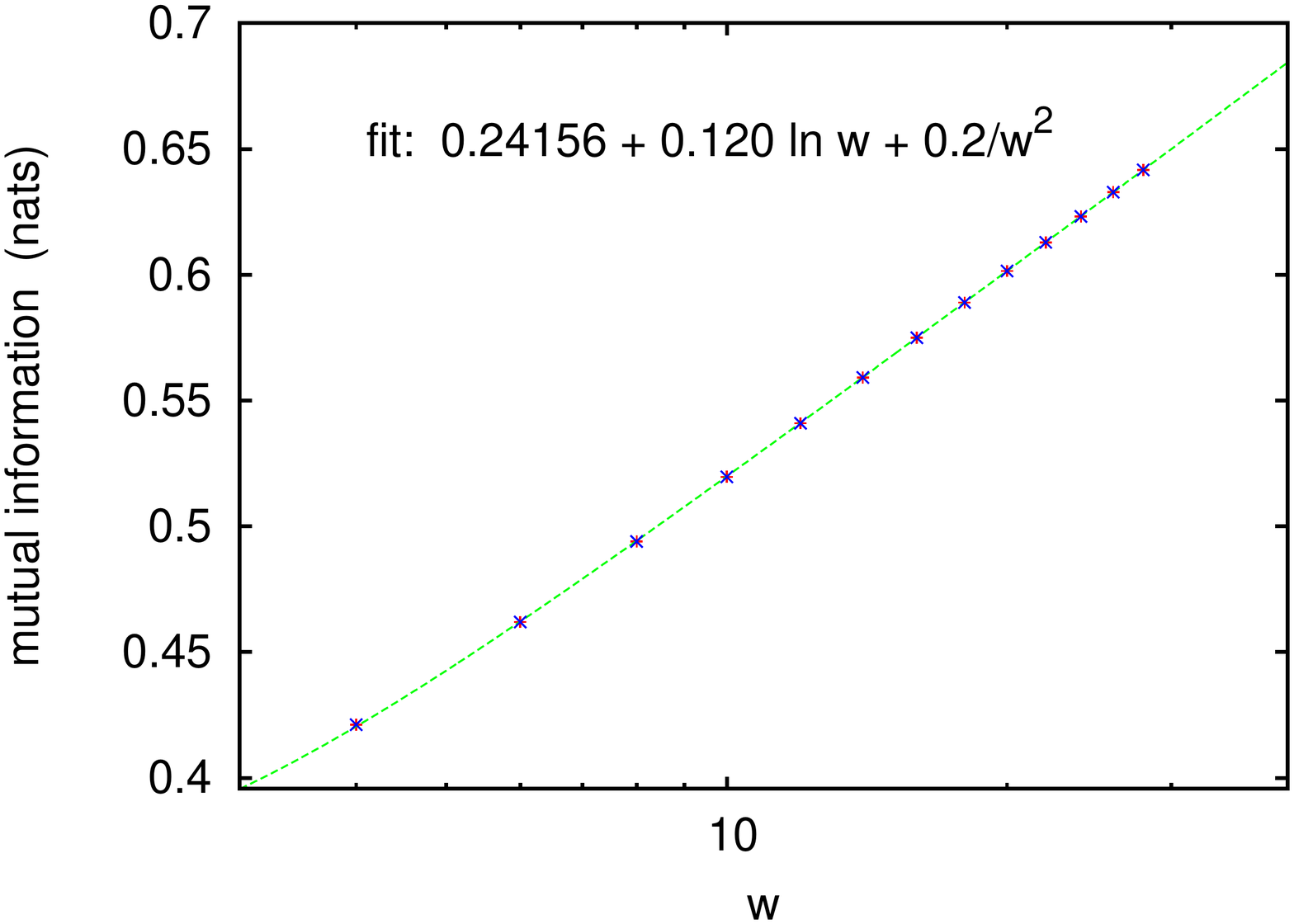}
\caption{(Color online) Mutual informations between two equal halves of a ring of $w=2m$ spins
(in natural units), plotted against $\log w$. The detailed form of the fit is presumably not to be 
relied upon, but the coefficient of the term linear in $\log w$ seems to be robust.}
 \label{ring-MI.fig}
\end{figure}

Data for $w=28$ at $\beta=\beta_c$ are shown in Fig.~\ref{ring-MI2.fig}. Together with the MI we show 
there a fit 
\be
   MI \approx a + b' \ln (\frac{w}{\pi} \sin\frac{m\pi}{w})  \label{ring-MI2-fit}
\ee
as suggested in \cite{Um-2012} for periodic chains of $w$ spins in the ground state of the quantum
Ising model in a transverse magnetic field (see also \cite{Calabrese-2004}). We see that the fit 
is not perfect (the points for $m=1$
and $m=w-1$ clearly deviate from it), but the overall agreement is surprisingly good. The constant $a$
is obtained as $a=0.380(1)$, which is clearly different from the value $0.329$ found in \cite{Um-2012},
suggesting that $a$ is not universal. In order to compare $b'$, we first plot the maximal MI (for $m=w/2$)
against $w$, see Fig.~\ref{ring-MI.fig}. We see clearly a logarithmic increase. Fitting the values for 
even and odd $w$ with correction terms $\propto 1/w^2$ or $1/w$ gives the estimate
\be
   b' = 0.1201(2) \;\;{\rm nats},
\ee
corresponding to $0.1733(3)$ bits. Within less than 1\% this agrees with the value found in \cite{Um-2012}.

Thus we conjecture that $b'$ is universal. It should be related to the universal constant $r_c$ discussed 
in the previous section, but the detailed relationship is not clear. Further questions of universality
are discussed in the next section.

\section{Entropies of loops and open-ended strings from Monte Carlo simulations}

\begin{figure}
\includegraphics[width=0.58\textwidth]{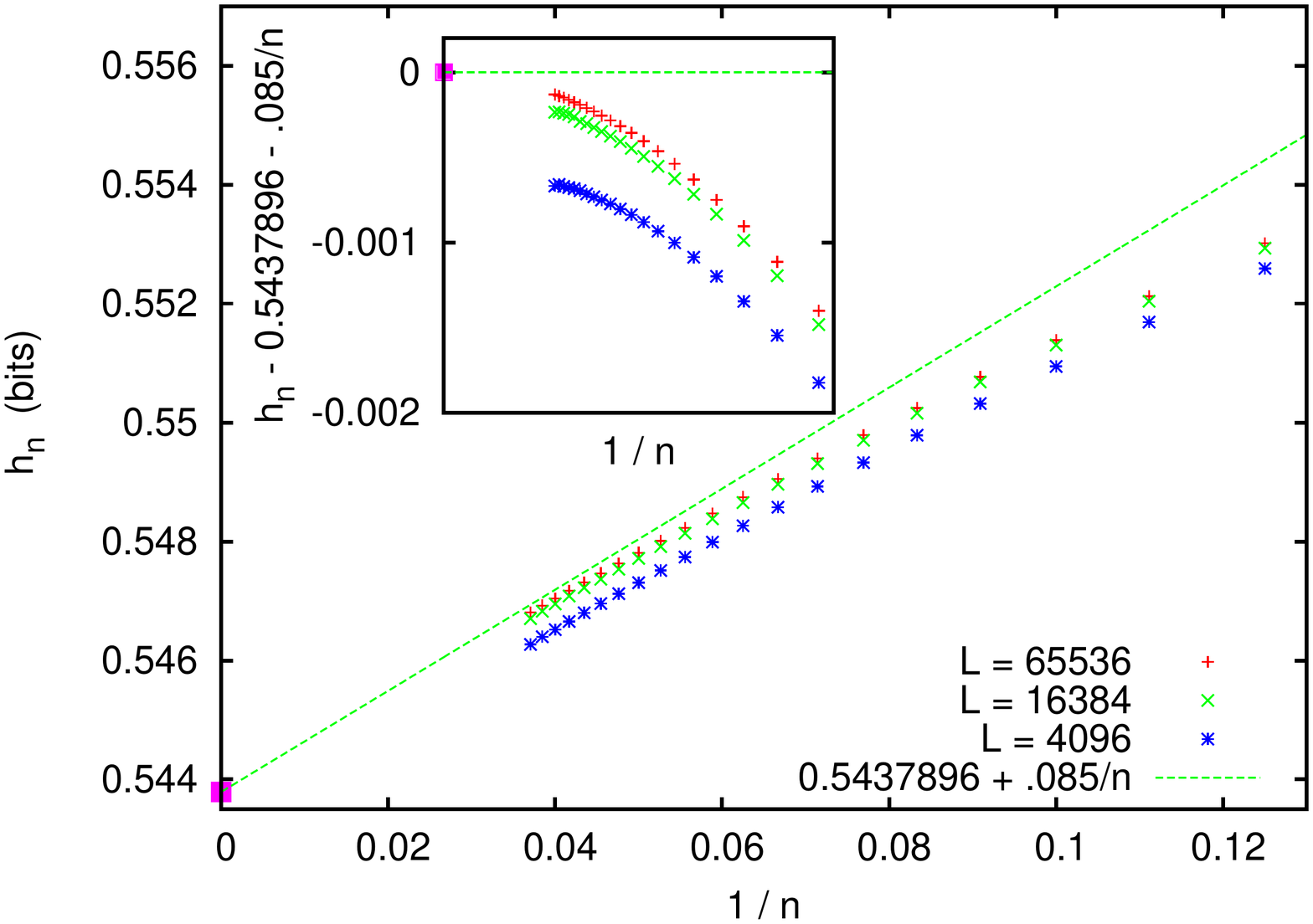}
\caption{(Color online) Convergence of $h_n$, the entropy per spin in a line of length $n$ embedded
in lattice of size $L\times L$ with helical (i.e. practically periodic) b.c. On the $x$-axis is plotted
$1/n$, so that a straight line indicates a convergence $h\approx h + b/n$. The straight line 
passes has $h=0.54378\ldots$ as obtained for rings encircling cylinders.The inset shows the data 
after subtraction of this linear part.}
 \label{Wolff-entropy.fig}
\end{figure}

Finally we wanted to see whether the entropy of a set of $n$ spins embedded in an infinite 
$2-d$ lattice shows any anomalous behavior at the critical point, when $n\to\infty$.

Let us look first at straight lines of spins. We expect of course that 
$h=\lim_{n\to\infty} h_n$ is the same as the entropy per site in an infinitely long ring encircling
a cylinder. But it is a priori not clear whether the excess entropy -- i.e. the MI between two halves 
of the line also diverges when $n\to\infty$ and $T\to T_c$.

Previous studies \cite{Arnold-1996,Kenneway-2005,Melchert-2012} have all found that the excess 
entropy ${\cal E} = {\cal E}({\cal S}_n)$ has a maximum near $T=T_c$ \footnote{For $T<T_c$ the value of 
${\cal E}$ depends on the ensemble considered. If one takes an ensemble with broken symmetry, e.g. with
positive magnetization, then ${\cal E} \to 0$ for $T\to 0$. Otherwise, if both phases are included, 
${\cal E} \to 1$ bit for zero temperature.},
but none did find any divergence. All these 
studies used Monte Carlo (MC) simulations. In view of the difficulties measuring ${\cal E}$ precisely in such 
simulations, this should not be taken as a real proof that ${\cal E}$ remains finite.

These difficulties and our strategies used to overcome them are detailed in Appendix C. The final
result is shown in Fig.~\ref{Wolff-entropy.fig}, where we plotted $h_n$ versus $1/n$, for three
lattice sizes ($L=4096, 16384$, and 65536). The straight line is such that it crosses the $y$-axis
at $h = 0.543789\ldots$ as determined in the previous section, and becomes tangent to the $L=65536$
data extrapolated to $1/n\to 0$. It has a slope of $b = 0.085(1)$. This means that 
\be
   H_n \approx nh + b \ln n\;\; {\rm bits},\quad b = 0.085\pm 0.001.    \label{log}
\ee

\begin{figure}
\includegraphics[width=0.5\textwidth]{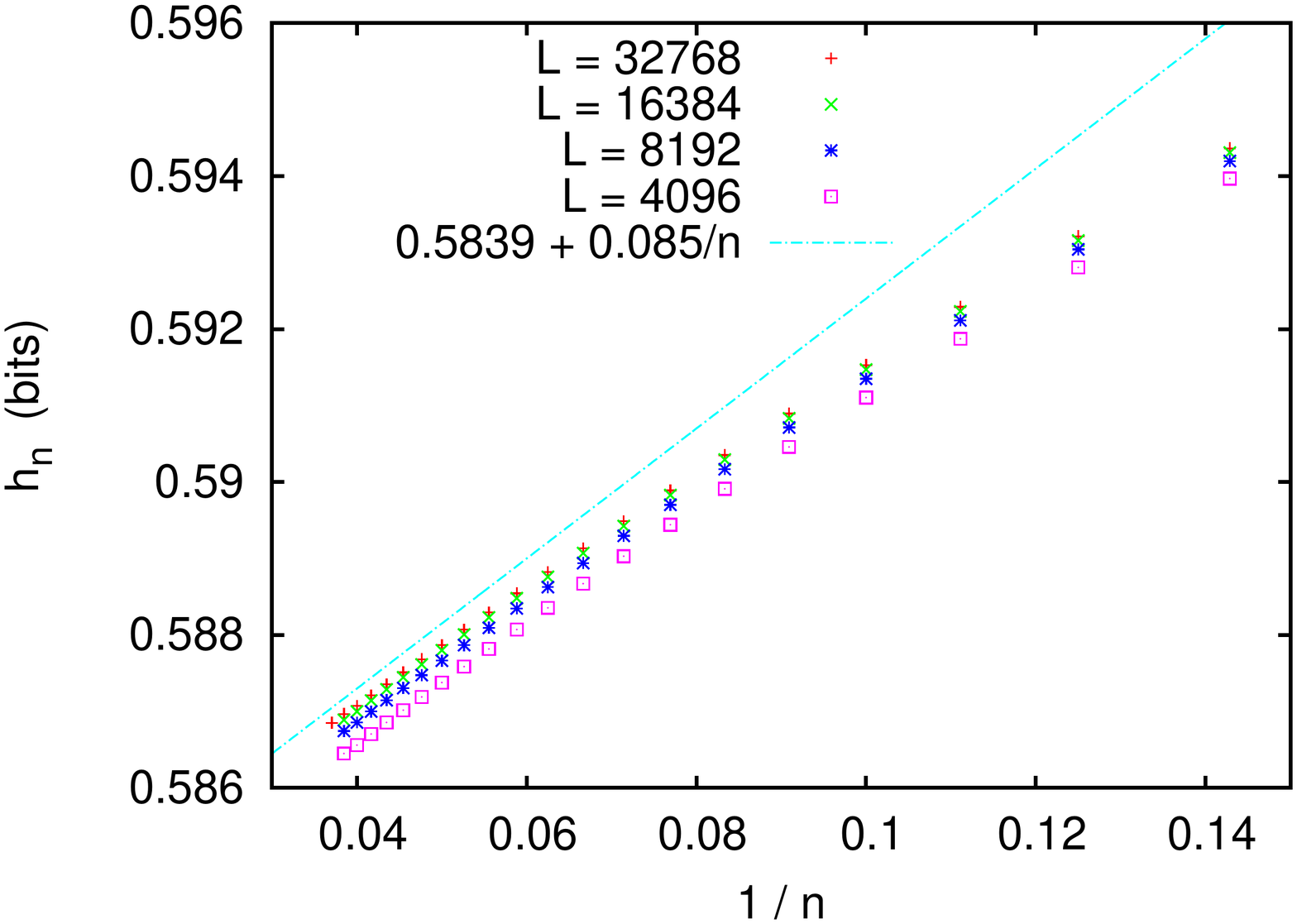}
\caption{(Color online) Convergence of $h_n$ for a line of spins embedded in a triangular 
lattice. In contrast to Fig.~\ref{Wolff-entropy.fig}, now the slope of the straight line fit is 
fixed (to the value $0.085$ obtained for the square lattice), but its intercept is fitted. 
From the latter we obtain $h=0.5839(3)$ for the triangular lattice.}
 \label{triangle-entropy.fig}
\end{figure}

\begin{figure}
\includegraphics[width=0.5\textwidth]{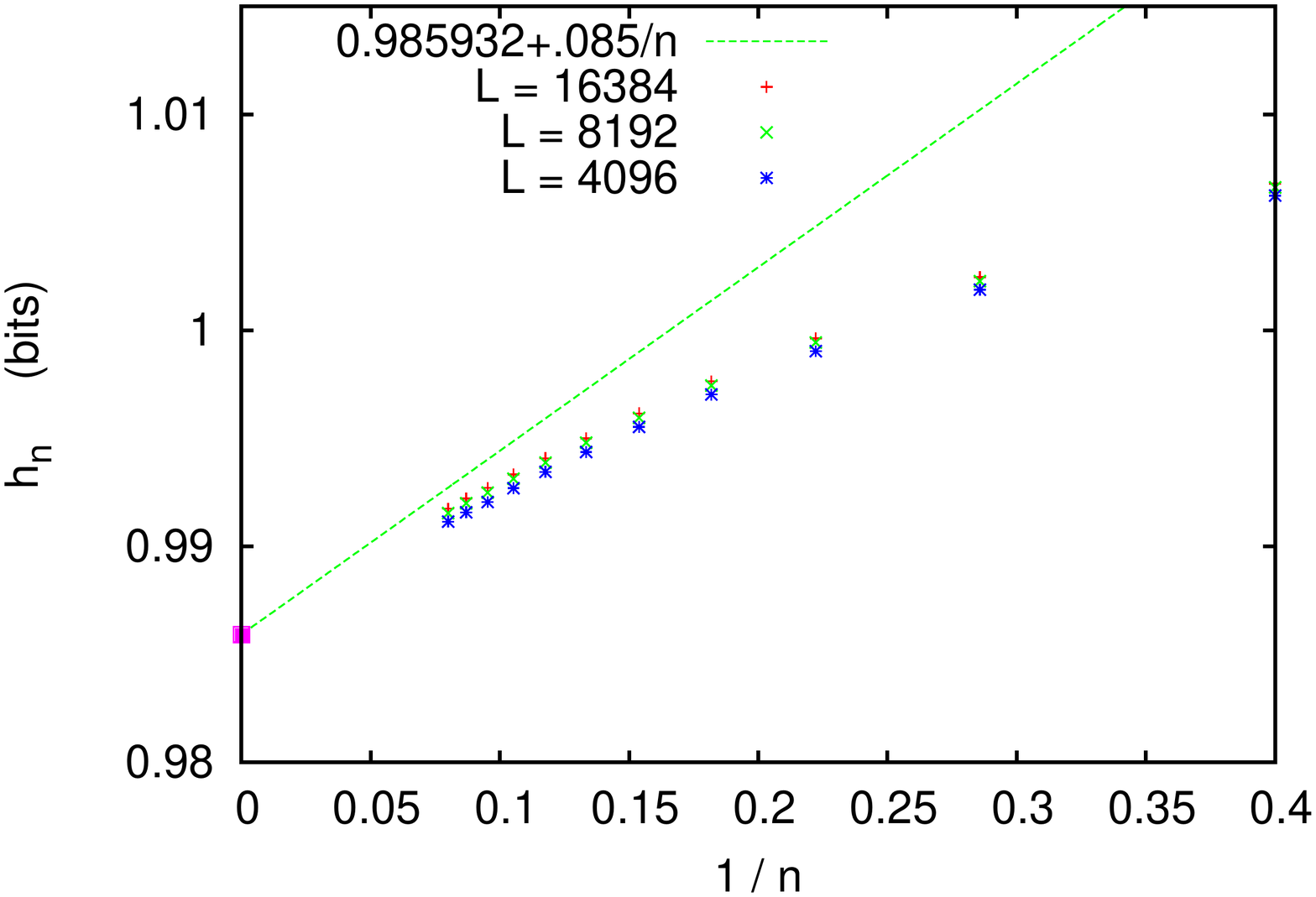}
\caption{(Color online) Convergence of $h_n$ for an $n\times 2$ rectangular block of spins 
embedded in a square
lattice. The straight line has the same slope as in Fig.~\ref{Wolff-entropy.fig}, and its intercept
is also given by the results of the previous sections. Thus it involves no new fitting parameter.}
 \label{2-strip.fig}
\end{figure}

\begin{figure}
\includegraphics[width=0.5\textwidth]{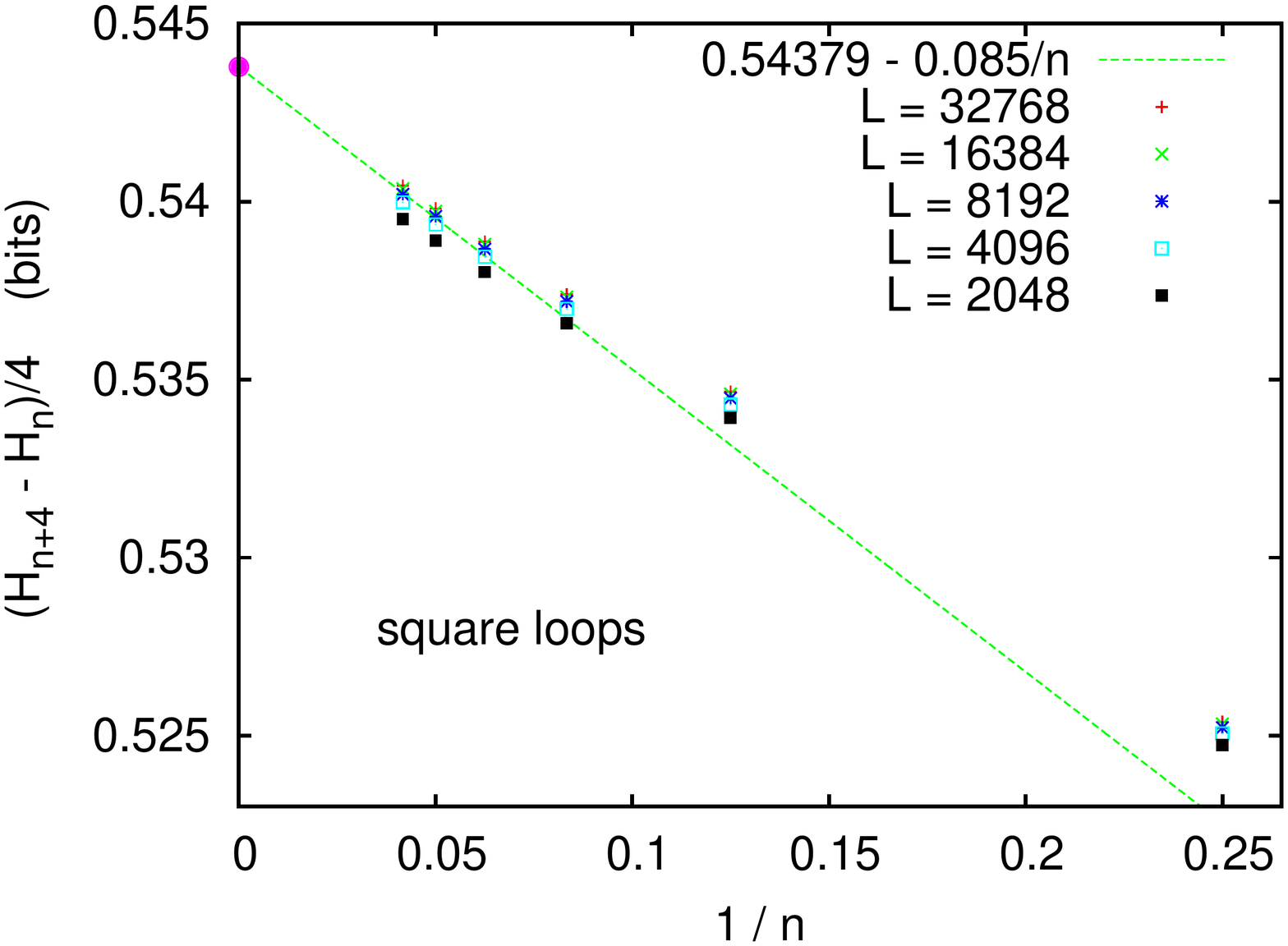}
\caption{(Color online) Analogous to Fig.~\ref{Wolff-entropy.fig}, but for square loops instead
of straight lines. Here, $n=4k$ is the length of the loop ($k=1,2,\ldots 6$). As in 
Fig.~\ref{Wolff-entropy.fig}, the intercept of the straight line is fixed, while its slope is 
fitted to the large-$n$ behavior for the largest value of $L$.}
 \label{square-loop.fig}
\end{figure}

When compared to the results of the previous section, we see that $b$ is very close to 
$b'/2$ (within one standard deviations). The relationship $b'=2b$ would be quite plausible,
given the fact that $\delta h_m$ is non-zero both for $m=O(1)$ and for $w-m=O(1)$. It can be 
proven exactly for entanglement entropies in quantum spin chains at $T=0$ \cite{Calabrese-2004}, 
but it is not clear whether the proof holds also in the present case. 

On the other hand, it seems natural to conjecture that $b$ is universal. To test this, we 
performed analogous simulations also for the triangular lattice. The Monte Carlo simulations 
used the same system sizes and had the same statistics. But, since we only 
wanted to check whether the data are also fitted by Eq.~(\ref{log}) with the same value of 
$b$, we did not perform any transfer matrix calculations. Thus the fit shown in 
Fig.~\ref{triangle-entropy.fig} is not constrained to pass through a precisely known value 
of $h$ for $1/n=0$, in contrast to that in Fig.~\ref{Wolff-entropy.fig}. Yet, the result clearly 
suggests universality as regards the type of lattice.

An indication that universality holds even more generally comes from looking at lines of width 2.
In Fig.~\ref{2-strip.fig} we show the entropies $h_n$ (in bits) when the ``alphabet" is a 
pair of spins $s^{[2]}$ (so the string is of the form ${\cal S}_n^{[2]}$), and $H_n = 
H_n({\cal S}_n^{[2]})$ is the entropy needed to specify the spin configuration on a 
$2\times n$ rectangular block. In the limit $n\to\infty$ this block becomes 1-dimensional, and 
\bea
   \lim_{n\to\infty} h_n &=& \lim_{w\to\infty}w^{-1}H_2(w,\beta_c) \\ 
                 & \approx & 0.442143 + 0.543790 = 0.985933 \;\;{\rm bits}.  \nonumber
\eea
Thus we can make a constraint fit as in Fig.~\ref{Wolff-entropy.fig}, the result of which 
is shown in Fig.~\ref{2-strip.fig}. We see that there are now much larger corrections to scaling
(as expected), but a decent fit (with no new parameters!) is obtained with the same $b$ 
as found in the above two cases.

Consider finally a set of $n$ spins, not forming necessarily a straight line.
An extended conjecture of universality would be that the value of $b$ depends only 
on the gross geometric shape of this set. As a first test that different shapes 
give rise to Eq.~(\ref{log}) but with different values of $b$, we considered loops of 
$n=4k$ spins on the square lattice formed by four straight legs of length $k$ each. Although 
corrections to scaling are again large, the data shown in Fig.~\ref{square-loop.fig} clearly 
suggest that Eq.~(\ref{log}) holds, with $b<0$ in this case.

\section{Conclusion}

In the present paper we have studied quantities related to Shannon entropies needed 
to specify the states of various sets of spins. A natural extension of this work would be 
the study of R\'enyi entropies, as e.g. in \cite{Stephan-2010}, another one to different 
models like the Potts or Blume-Emery-Griffiths models \cite{Blume}. 

Some of the entropy measures studied (like, e.g., the average entropy $h_c$
given in Eq.~(\ref{hw})) are related to the thermodynamic entropy, but not all. The reason is
on the one hand that we studied several MIs that have no direct counterpart 
in thermodynamics. On the other hand, every Shannon information has to be understood relative
to some conditioning, and these conditionings differ for the different entropies studied 
above. The very concept of MI is closely related to this, as the MI 
between ${\cal A}$ and ${\cal B}$ is just the difference between the unconditional 
information needed to specify ${\cal A}$ and the information needed to specify ${\cal A}$, 
conditioned on knowing ${\cal B}$. 

In cases where one wants to describe the state of a bi-infinitely long string of ``letters",
the MI between the two halves of the string is called its {\it excess entropy}. 
This is the quantity most directly involved, if we want to know whether the entropy is strictly 
extensive or deviates from extensivity due to long range correlations. The 2-dimensional 
Ising model was studied because it develops such long range correlations exactly at the critical
point. It is thus natural to expect strong corrections to extensivity exactly at the critical
point. We showed that this is indeed the case: The information needed to describe the states 
of contiguous strings of $n$ spins contains in general a term $\propto \ln n$. The amplitude 
in front of this term depends on the geometry of the string in the large $n$ limit (it seems,
e.g., to be twice as large for rings encircling infinitely long cylinders than for open
strings), but it seems to be otherwise universal. 
Within numerics, it is the same for square and triangular lattices, and for blocks of size
$1\times n$ and $2\times n$. We conjecture that such logarithmic corrections
hold for any family of subsets of spin that scales in the 
limit $n\to\infty$, and for any 2-dimensional critical phenomenon -- or maybe even in higher
dimensions. We suggest that this is the main open question raised by the present study.

On the other hand, we showed explicitly that long range correlations alone do not by necessity 
lead to large excess entropies and thus to strong deviations from extensivity. Even when the 
correlations diverge at the critical point, the Ising model is still Markovian, as best seen
from the Markovian structure of the transfer matrix. If all relevant degrees of freedom are 
explicit, then all MIs between two regions are bounded by the entropy of the 
interface between them. In the case of a long strip of finite width, the excess entropy is 
thus bounded by the width and cannot diverge at the critical point. It is only when one 
considers long strings of spins embedded in a large background which is {\it not} treated
explicitly, that diverging mutual entropies can occur. 

This remark is also relevant for the holographic principle in classical spin systems 
\cite{Wolf-2008}. In contrast to the original holographic principle for black holes, where 
the {\it entropy} is given by the surrounding area \cite{Bousso-2003}, the holographic principle
in statistical mechanics stipulates that the {\it mutual information} between a finite 
(sub-)system and its environment is bounded by the interface area. In general one might suspect
that this interface ``gets fuzzy" at a critical point, and that its area should therefore be 
replaced by the product between the area and the correlation length \cite{Wolf-2008}. In several 
cases it was found that this is not needed, and the reason is obvious from the above: The 
relevant ``thickness" of the interface is not given by the correlation length, but by the 
order of the Markov field -- which is small for most models studied in statistical 
physics.

Although we have studied only 2-dimensional systems in the present paper, we expect most 
results to carry over to higher dimensions. In particular, the excess entropy for a long 
system with finite cross section should be finite, while the one for a string of spins
embedded in an infinite lattice should diverge logarithmically at the critical point.

Let us make a last remark on the logarithmic divergence of the excess entropy for spin chains. 
Superficially, this is very similar to the behavior of self-avoiding walks (SAW) \cite{deGennes}. 
For SAW in $<4$ dimensions the partition sum (i.e. the number of distinct configurations
of $n$-step walks) is given asymptotically by $Z_n \sim e^{\mu n} n^{\gamma-1}$. Thus the entropy 
(the logarithm of the partition sum) contains an extensive part $\mu n$ and a logarithmically
diverging part $(\gamma-1)\ln n$. The latter is universal with respect to the type of lattice, 
but depends on the topology of the SAW. In particular, $\gamma$ is different for open SAWs
and for closed loops \cite{deGennes}. Notice, however, an important difference to our present 
problem: While $\gamma$ is defined only by averaging over all walk geometries with a given
topology, the analogous constant $b$ defined in Eq.~(\ref{log}) is defined for each 
individual geometry.

\section{Appendix A}

Let us call the $2^w\times 2^w$ transfer matrix {\bf T}. Its leading eigenvalue $\lambda$ is 
related to the partition sum by $Z_L \sim \lambda^L$. The corresponding right and left 
eigenvectors $|\psi\rangle$ and $\langle \phi|$ are normalized such that $\langle \phi|\psi\rangle=1$.
It is well known that the probability for the spin state $i \in \{-1,1\}^w$ is given by
\be
   p_i = \langle \phi|i\rangle\langle i|\psi\rangle = \psi_i \phi_i. 
\ee
Our strategy for calculating $H_1^w$ is thus to iterate simultaneously the two equations
\be
   \psi_{i,t+1} = \lambda^{-1} \sum_j T_{ij}\psi_{j,t}\;,\quad \phi_{i,t+1} = \lambda^{-1} \sum_j T_{ji}\phi_{j,t}
            \label{T_iter}
\ee
and normalize them, until 
\be
   H_{1,t}^w = - \sum_i p_{i,t} \log p_{i,t} \;\quad {\rm with}\;\;\; p_{i,t} = \psi_{i,t} \phi_{i,t}
\ee
has converged to its limit $H_1^w = \lim_{t\to\infty} H_{1,t}^w$. 

Actually, we did not use in Eqs.(\ref{T_iter}) the transfer matrix for adding an entire line, but 
we wrote {\bf T} as a product over $w$ `partial' transfer matrices where we added one spin in each 
\cite{Morgenstern-1979}.
The advantage is that these partial transfer matrices are sparse (they have just two entries in each 
row), and thus the CPU time required for one iteration is reduced from ${\cal O}(4^w)$ to 
${\cal O}(4w2^w)$ (one looses a factor 2 because the partial transfer matrices are not symmetric, 
whence both iterations of Eqs.(\ref{T_iter}) have to be actually done, while they would be identical
if the full transfer matrix were used). This allowed us to obtain $H_1^{w=29}$ in 35 CPU hours on a 
modern workstation. All calculations were done in extended (80 bits) precision, and it was checked that 
they agreed with results obtained with 64 bit double precision.

\section{Appendix B}

In \cite{Stephan-2010}, the value of $r_c = r(\beta_c)$ was obtained by extrapolating data for the 1-d
Ising chain in a transverse field, for $w=16, 18, \ldots 44$. While the raw data (i.e. the values of
$H_1(w,\beta_c)$) are precise up to 15 digits, we believe that the extrapolation for $w\to\infty$ was
not done optimally. We thus present here an alternative analysis that provides, in out opinion, a much
more precise value of $r(\beta_c)$.

\begin{figure}
\includegraphics[width=0.5\textwidth]{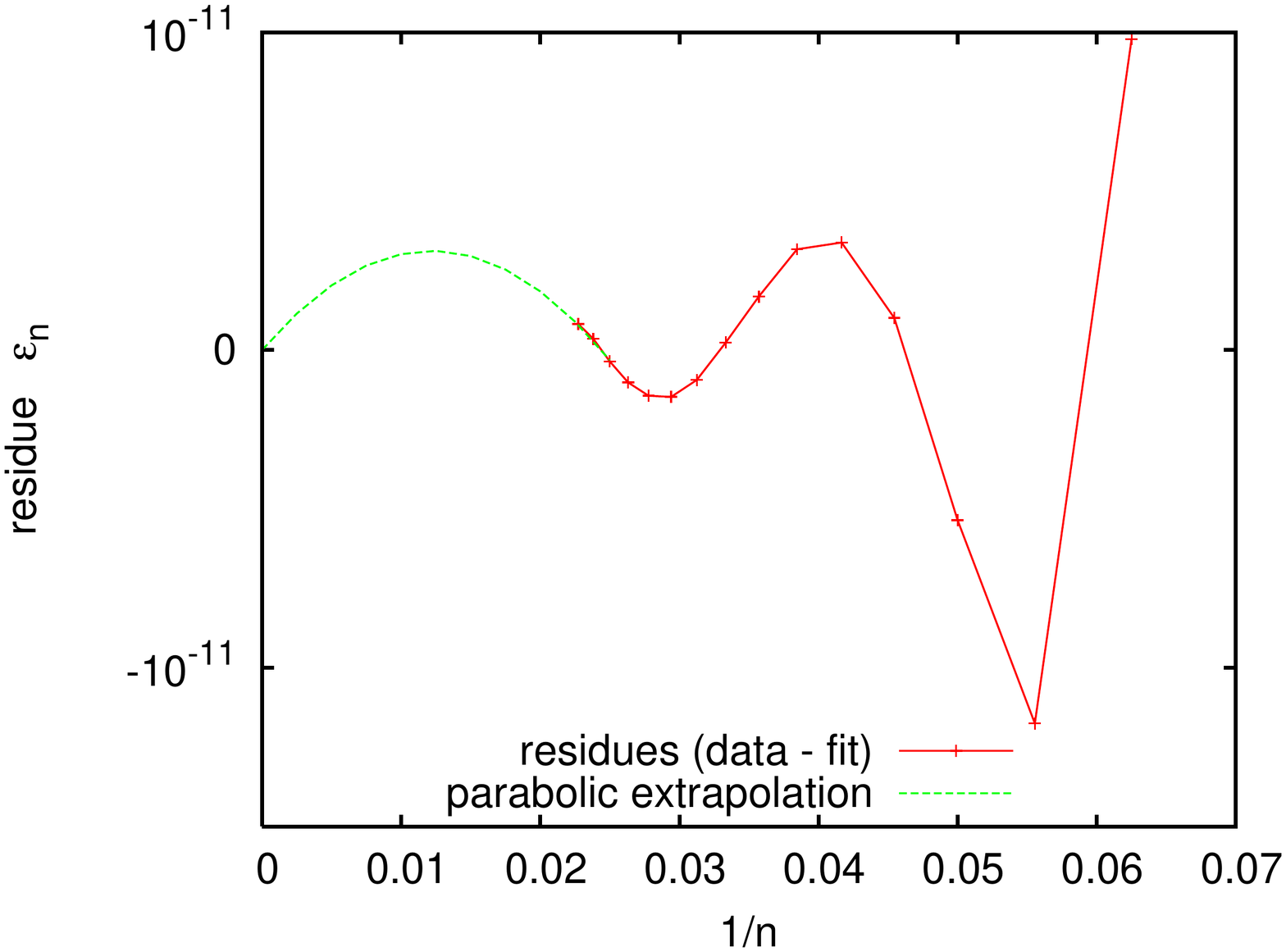}
\caption{(Color online) Residues in fitting $w^{-1}H_1(w,\beta_c) - f(w)$ where the fit $f(w)$ is a
    polynomial in $1/w$ (see Eq.~(\ref{B0})),
    for the 1-d Ising chain in a transverse magnetic field, using the data from \cite{Stephan-2010}.
    Data are plotted against $1/w$. Thus, if both the data and the fit have simple asymptotics, the 
    residues should be smooth for small $1/w$ and their extrapolation should pass through zero. The 
    continuous curve is such an extrapolation. }
 \label{residues.fig}
\end{figure}

As in \cite{Stephan-2010}, our analysis is based on least-square fitting $w^{-1}H_1(w,\beta_c)$) by 
a polynomial in $1/w$,
\be
   w^{-1}H_1(w,\beta_c) \approx f(w):= a_0 + a_1/w + \ldots a_k/w^k,      \label{B0}
\ee
but the details of the fits are quite different.

At a first try, we fitted all 15 values (for even $w=16, \ldots 44$) with a polynomial of 
order 7. In order to obtain a good fit for large values of $w$, eventually at the cost of
obtaining a bad fit for small $w$, we made a weighted fit, i.e. we minimized the weighted sum
\be
   \chi_0 = \sum_w w^m [w^{-1}H_1(w,\beta_c) - f(w)]^2
\ee
with a large positive value of $m$ (in most fits we used $m=5$ to 7). Since the coefficients 
$a_k$ are not well constrained by such a fit and would e.g. increase rapidly with $k$, if there 
were non-analytic terms in the true $f(w)$, we also added to $\chi_0$ a penalty term of the form 
$\lambda \sum_k a_k^2$. Finally, we want $f(w)$
to be such that $f(w)\to h(\beta_c)$ for $1/w\to 0$, which means that the residues
\be
   \epsilon_w = w^{-1}H_1(w,\beta_c) - f(w)
\ee
should vanish for $w=0$. We cannot impose this as a constraint, of course, since we do not know 
the values of $w^{-1}H_1(w,\beta_c)$ for large $w$. But we know that both $w^{-1}H_1(w,\beta_c)$ 
and $f(w)$ should be smooth functions, and thus their difference (i.e. the residues) should be 
also smooth. We thus added yet another term that controls the derivative of $f(w)$ at the largest 
accessible value of $w$. The total cost function to be minimized is thus
\be
   \chi = \chi_0 + \lambda \sum_k a_k^2 + \mu [\epsilon_{44}-\epsilon_{42}-b]^2,   \label{B1}
\ee
with $m,\lambda,\mu$ and $b$ free parameters.

After extensive trials, we found that:
\begin{itemize}
\item There is indeed no indication that $ w^{-1}H_1(w,\beta_c)$ is not described by a power 
series in $1/w$, and even without any penalty the coefficients $a_k$ tend very rapidly to zero 
for large $k$. We thus put $\lambda=0$ and we truncated $f(w)$ to a fifth order polynomial. Higher 
order terms would have very small coefficients and have virtually no effect on the estimates of 
$r_c$ and $h(\beta_c)$.
\item The power $m$ can be fixed to $m=6$. Other values gave very similar results.
\item The last (slope-controlling) term in Eq.~(\ref{B1}) is crucial. We fixed $\mu$ and $b$ such 
that an extrapolation of the residues quadratic in $1/w$ would pass through $\epsilon_\infty = 0$
(see Fig.~\ref{residues.fig}).
\end{itemize}

Residues obtained by a typical fit, together with their extrapolation to $1/w\to 0$, are shown in 
Fig.~\ref{residues.fig}. For all $w > 18$ they are $<10^{-11}$. By comparison, the fits used 
in \cite{Stephan-2010} typically had residues $\approx 10^{-9}$. The corresponding values of 
$r_c$ and $h(\beta_c)$ are given in Eqs.~(\ref{rc},\ref{hc}). The
errors are obtained by trying different constants in Eq.~(\ref{B1}).

Once $r_c$ has been obtained in this way, we can use it as a constraint when estimating 
$h(\beta_c)$ for the classical Ising model. In that case we found again no hint for a non-analytic 
behavior in $1/w$. On the other hand, the coefficients $a_k$ did not decrease with $k$ as fast as 
for the 1-d Ising chain in a transverse field, whence we used an eighth order polynomial and we 
used a small positive value of $\lambda$ to keep the coefficients small.
Again the last term in Eq.~(\ref{B1}) was important.

\section{Appendix C}

For simulating the Ising model, we used the Wolff \cite{Wolff} algorithm. When estimating entropies 
with high precision from histograms generated by Monte Carlo simulations, one has to cope with three 
problems: Finite size corrections, transients, and finite sample corrections in the histograms.

First simulations on a lattice of size $2048\times 2048$ suggested important finite size corrections, 
whence we finally made simulations on lattices with $L=4096, 16384$, and $65536$. As seen from 
Fig.~\ref{Wolff-entropy.fig}, the latter is indeed needed, but is also big enough. We used helical 
boundary conditions.

\begin{figure}
\includegraphics[width=0.5\textwidth]{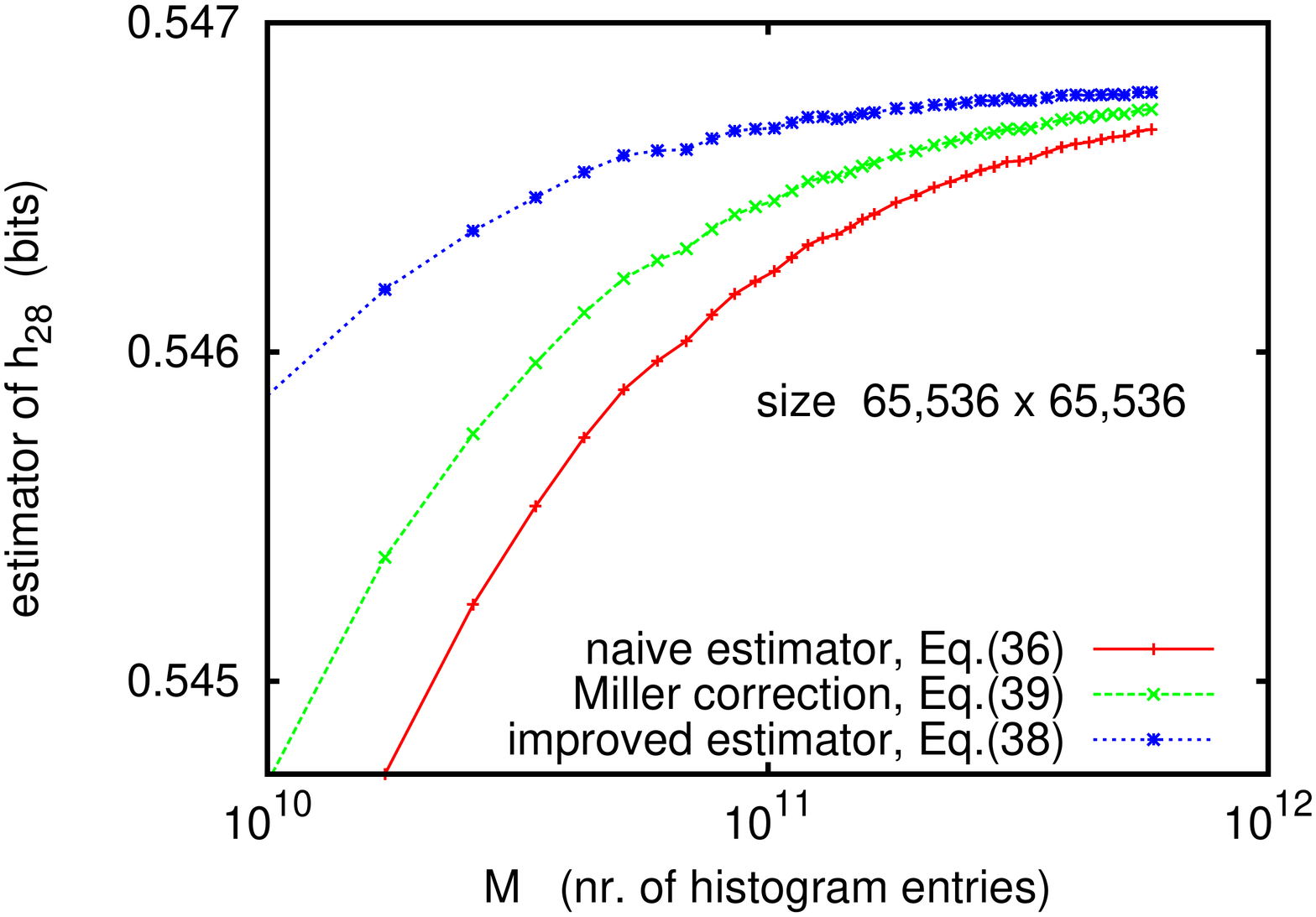}
\caption{(Color online) Convergence of entropy estimators for the critical 2-d Ising model. }
   \label{fig-AC1.fig}
\end{figure}

As concerns transient times, we started off using random initial configurations. In that case, 
our results at $T_c$ were clean only when we discarded transients 
with $\gg 10^3$ spin flips per site. 
This might seem to contradict the supposed very short correlation time for the Wolff algorithm, 
but the latter is only for correlations in equilibrium, not for the relaxation towards it. If one 
starts with a random spin configuration, transients are dominated for a long time by 
configurations where part of the lattice has already long range correlations, while the other 
part is still disordered. During this time the number of flipped {\it clusters}
is roughly the same in both regions, but the number of {\it spin} flips is vastly larger 
in the former. Thus it takes very long -- when time is measured in terms of spin flips -- until 
the latter also gets ordered. This effect is decreased by starting with a finite (but not too 
large) magnetization. We found empirically that transients were shortest when the starting
configuration was at the percolation threshold, i.e. when one of the first clusters to be flipped 
was a giant cluster spanning the entire lattice.

After the transient, we made histograms with $2^n$ entries with $n=29$.
Each entry corresponds to a horizontal or vertical line $S^n$ of spins. This 
histogram was updated after every $\approx 50$ spin flips per site, by shifting $S^n$ through 
all $L^2$ positions. The simulations were stopped when the histograms had $M> 5\times 10^{11}$ 
entries, corresponding to about six days of CPU time. From any histogram with $2^n$ entries 
we can obtain histograms with $2^{n'}$ entries with $n'<n$ corresponding to shorter strings by 
coarse graining. 

In spite of the very large number $M$ of histogram entries, there are still substantial 
finite-$M$ corrections to the naive entropy estimator
\be 
   \hat{H}_n = -\sum_{i=1}^{2^n} \frac{m_i}{M} \log\frac{m_i}{M}                  \label{naive}
\ee
where $m_i$ is the number of entries in the $i$-th slot of the histogram, with $\sum_i m_i = M$.
More precisely, this estimator always underestimates the entropy, by mistaking statistical 
fluctuations for real (entropy-reducing) structure.
Therefore we used the improved estimator from \cite{Grass-2003}. In this estimator the logarithm is 
first split up into $\log m_i - \log M$, and then the (natural) logarithms of integers are replaced
by a function $G_m$, where 
\be
   G_0 = G_1 = -\gamma - \ln 2, \quad G_{2n+2} = G_{2n}+\frac{2}{2n+1},    \label{G}
\ee
and $\gamma=0.577215\ldots$ is the Euler-Mascheroni constant, giving
\be
   \hat{H}_{G,n} = G_M -M^{-1}\sum_{i=1}^{2^n} m_i G_{m_i} \;.                 \label{HG}
\ee
Other estimators with smaller 
bias do exist \cite{Grass-2003}, but there is in general a trade-off: One can either minimize the 
bias or the statistical variance of any entropy estimator, but not both. Eq.~(\ref{G}) is constructed 
such that it is close to optimal in the case where $m_i \ll M$ for all $i$. The estimates for the 
block entropy $h_{28} = H_{29}-H_{28}$ for a $65536^2$ lattice are shown in Fig.~\ref{fig-AC1.fig}.
We see there that the naive estimator would still be substantially biased, while both the bias 
and the statistical error for the improved estimator are $\approx 10^{-5}$. In Fig.~\ref{fig-AC1.fig}
we also show the naive estimator with Miller correction \cite{Miller},
\be
   \hat{H}_{{\rm Miller},n} = \hat{H}_n + \frac{N_n-1}{2M},    \label{Miller}
\ee 
where $N_n$ is the number of non-zero 
histogram entries. Although the Miller correction removes the bias to leading order in the limit 
$M\to\infty$, $\hat{H}_{{\rm Miller},n}-\hat{H}_n = o(1/M)$, it corrects only for about half of the bias 
in the present case.

\section*{Acknowledgements}

We thank Deepak Dhar for discussions and for carefully reading the manuscipt. We also thank Johannes Wilms
for correspondence which actually triggered this research.

\bibliography{mm}

\end{document}